\documentclass[aps,prl,twocolumn,showpacs,superscriptaddress]{revtex4-2}

\usepackage{amsmath}
\usepackage{graphicx}
\usepackage{lmodern}
\usepackage{amsmath}

\usepackage{color}
\usepackage{amssymb}
\usepackage{bm}
\usepackage{braket}
\usepackage{mathtools}

\newcommand{\up}{\uparrow}
\newcommand{\down}{\downarrow}
\newcommand{\mr}{moir\'{e} }
\newcommand{\BZ}{Brillouin zone }
\newcommand{\tr}{\text{Tr}}

\usepackage[dvipsnames]{xcolor}
\usepackage[colorlinks=true]{hyperref}
\hypersetup{
    colorlinks=true,
    linkcolor=blue,
    citecolor=blue,
    urlcolor=blue
} 

\makeatletter
\def\maketitle{
\@author@finish
\title@column\titleblock@produce
\suppressfloats[t]}
\makeatother

\begin{document}

\title {Pair-density-wave and chiral superconductivity in twisted bilayer transition-metal-dichalcogenides}
\author{Yi-Ming Wu}
\thanks{These authors contributed equally to the work.}
\affiliation{Institute for Advanced Study, Tsinghua University, Beijing, China}
\author{Zhengzhi Wu}
\thanks{These authors contributed equally to the work.}
\affiliation{Institute for Advanced Study, Tsinghua University, Beijing, China}
\author{Hong Yao}
\email{yaohong@tsinghua.edu.cn}
\affiliation{Institute for Advanced Study, Tsinghua University, Beijing, China}
\affiliation{State Key Laboratory of Low Dimensional Quantum Physics, Tsinghua University, Beijing 100084, China}
\date{\today}

\begin{abstract}
We theoretically explore possible orders induced by weak repulsive interactions in twisted bilayer TMD (e.g. WSe$_2$) in the presence of an out-of-plane electric field. Using renormalization group (RG) analysis, we show that superconductivity (SC) survives even with the conventional van Hove singularities.  We find topological chiral superconducting states with Chern number $\mathcal{N}=1,2,4$ (namely $p+ip$, $d+id$, $g+ig$) appear over a large parameter region with \mr filling factor around $n=1$. At some special values of applied electric field and in the presence of a weak out-of-plane Zeeman field, spin-polarized pair density wave (PDW) superconductivity can emerge. This spin-polarized PDW state can be probed by experiments such as spin-polarized STM measuring spin-resolved pairing gap and quasi-particle interference. Moreover, the spin-polarized PDW could lead to spin-polarized superconducting diode effect. 

\end{abstract}
\maketitle
{\bf\large\noindent Introduction}

The advent of engineering
\mr structures
from stacking 2D materials starts an exciting path for studying intriguing electronic properties in quantum materials.  
Through interlayer van der Waals coupling, spatial \mr potential profoundly modifies electronic band structures, and in certain circumstances results in low energy isolated narrow or even flat bands (\mr bands) \cite{Lopes2007,Morell2010,Bistritzer12233,Trambly2012,Moon2012,Lopes2012,Tarnopolsky2019,Po2018,Zou2018,Wolf2019,Koshino2018,Kang2018}, where interactions play a vital role in low energy physics.
From experimental aspect, various phases have been found in magic-angle twisted bilayer graphene (TBG), including correlated insulators \cite{Cao2018b,Xie2019,Jiang2019,Choi2019,Kerelsky2019}, superconductors \cite{Cao2018b,Matthew2019,Lu2019,Arora2020}, strange metal\cite{Polshyn2019,Cao2020a,Lyustrange}, magnetic phases \cite{Aaron2019,Li2020,ZhangYu2020,Saito2021} and quantum anomalous Hall states \cite{Serlin2020,tseng2022anomalous}.  Soon after this, similar phases are also found in other forms of \mr structures including twisted double bilayer and trilayer graphene systems \cite{Liu2020,Shen2020,Chen2019a,Chen2019b,Chen2020,Cao2020b,He2021,Park2021,Zhu2020}.  

\begin{figure}[t]
  \includegraphics[width=8.5cm]{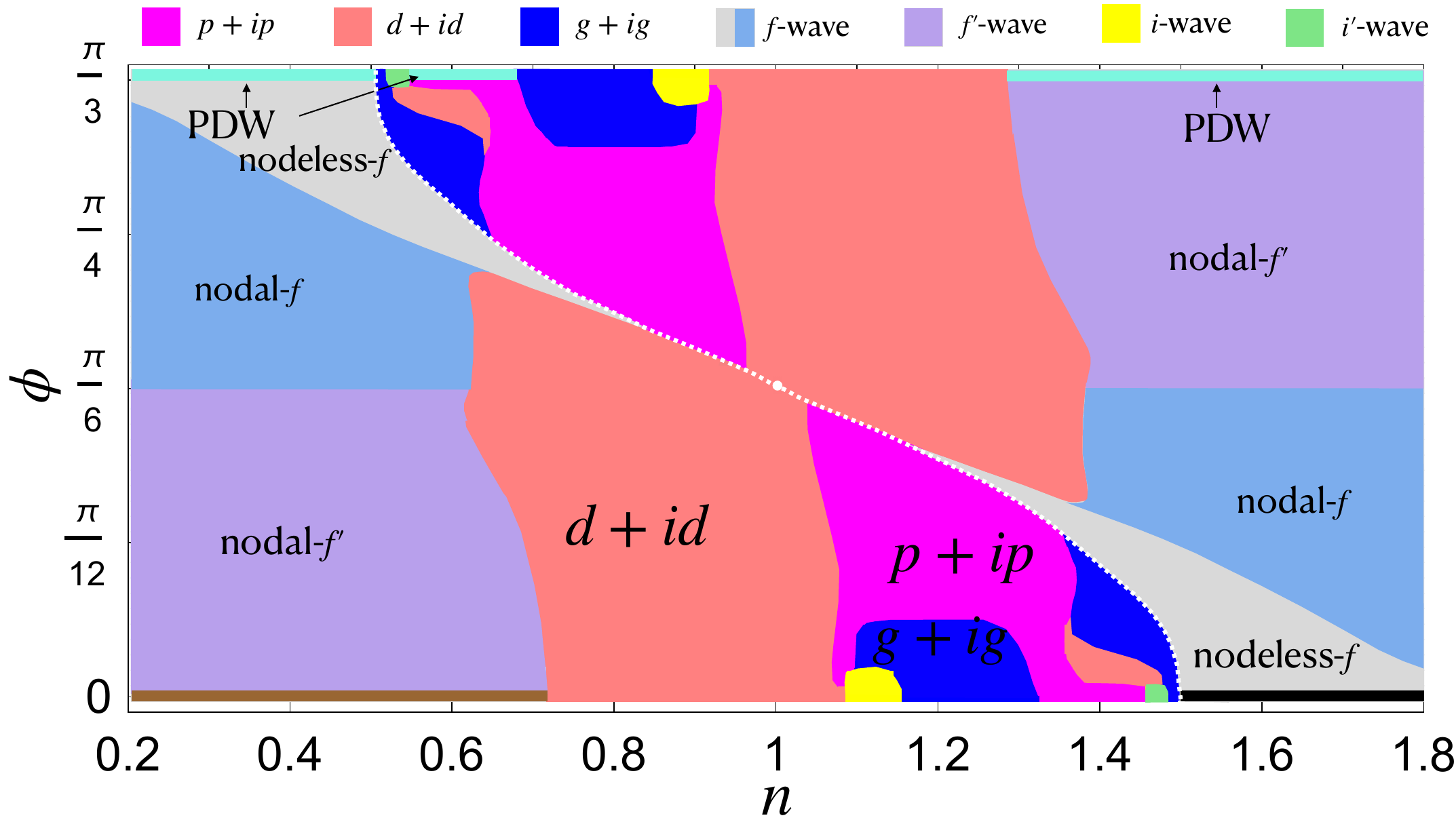}
  \caption{Phase diagram for twisted WSe$_2$. At $\phi=\pi/3$ there is an emergent spin-polarized PDW with $\bm{Q}=\pm\bm{K}$ at small and large $n$. Centered around $n=1$ there exist different chiral SC states which we identify as $p+ip$, $d+id$ and $g+ig$ based on their Chern numbers. At large or small $n$, a six node or nodeless SC state dominates, which we denote by $f$- or $f'$-wave according to their representation [$A_1$ ($B_{1u}$) or $A_2$ ($B_{2u}$) of $C_{3v}$ ($D_{6h}$) when $\phi>0$ ($\phi=0$)].  At $\phi\to0,\pi/3$, there are tiny regimes of 12 node SC states ($i$- and $i'$-wave differ in their representations). The white dashed curve shows the locations of the CVHSs. $\phi=\pi/6$ and $n=1$ is the location of the HOVHS, where the ground state is a gapless metal.}\label{fig:phase}
\end{figure}

Twisted bilayer transition metal dichalcogenides (TMD) 
were recently shown to be another promising and sometimes advantageous platform for simulating various correlated and topological 
states 
\cite{Wufc2018,Wufc2019,Zhang2020,Shabani2021,Weston2020,Devakul2021,ZhangYang2021,Angelie2021826118,Tran_2020,Vitale_2021,Bi2021,PhysRevB.102.235423}. Monolayer TMD is a semiconductor with strong spin-orbit coupling \cite{Manzeli2017,He2014,Jones2013,Srivastava2015,CHOI2017116,Mak2016}. 
The spin splitting in its valence band is much larger than that in the conduction band so that the topmost valence band can be used as a model for effectively spinless or spin-polarized fermions on Fermi surface \cite{Xiao2012}. Moreover, the finite energy gap in monolayer TMD enables the \mr band width to vary continuously with the twisted angle, implying a great tunability of \mr flat bands in this system compared to TBG \cite{Zhan2020}.
Very recently, correlated insulators, antiferromagnetism, and stripe phase have been observed in WSe$_2$/WS$_2$ heterostructure \cite{Tang2020,Regan2020,Xu2020,Jin2021,Huang2021}, where the \mr lattice is formed by the mismatch of the lattice constants. In MoTe$_2$/WSe$_2$ \mr heterostructures, metal-insulator transition 
was realized \cite{Li2021}. For twisted bilayer WSe$_2$ (tWSe$_2$), many interesting phenomena including metal-insulator transition, quantum critical behavior, and possible superconductivity (SC) were also found \cite{Wang2020,Ghiotto2021}.

\begin{figure*}[t]
  \includegraphics[width=15cm]{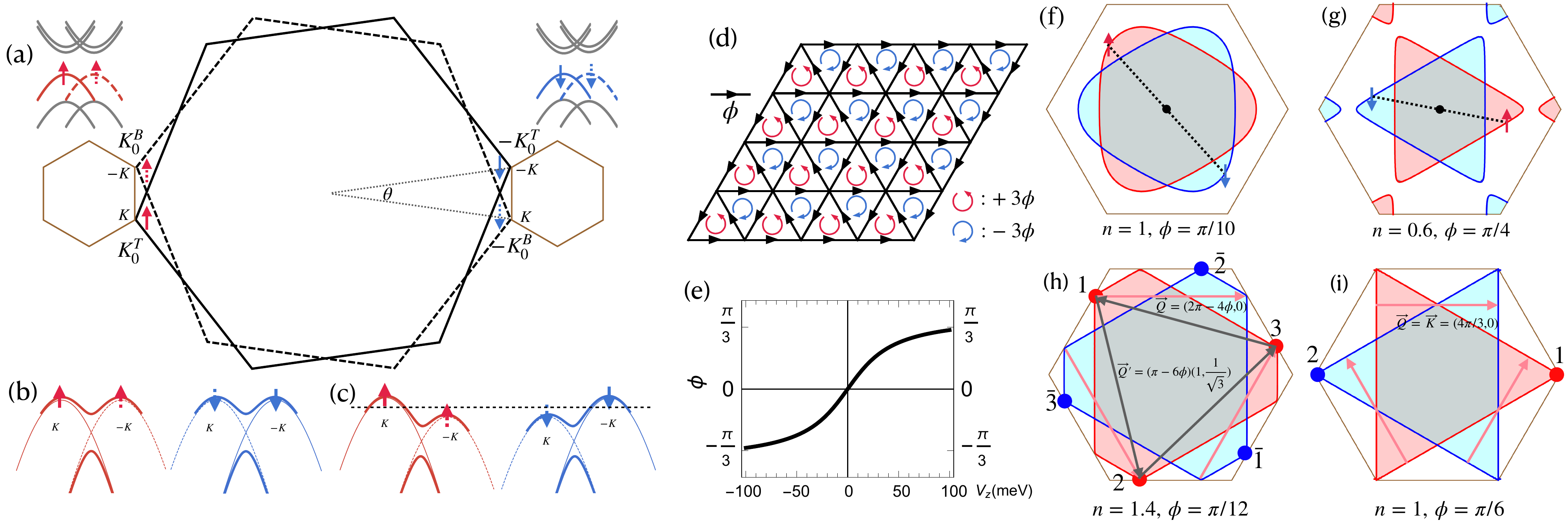}
  \caption{(a) Formation of the \mr \BZ in twisted homo-bilayer WSe$_2$. In the upper left and right insets we show two schematic plots for both the monolayer conduction and valence  band structures near $\pm K_0$. (b) Two fold degeneracy of the topmost \mr valence band, which can be lifted by an out-of-plane electric field (c). (d) The electric field effectively induces a stagger phase factor $\pm\phi$ on the bonds, leading to an accumulated flux $\pm3\phi$ on each triangular plaquette. (e) The physical bond phase $\phi$ as a function of $V_z$, from Ref.\cite{Pan2020}. (f-i) Various types of Fermi surfaces in the first \BZ. Depending on the parameters, there can be electron or hole pockets, disjoint Fermi surfaces, six CVHSs and two HOVHSs. The arrows in the last two figures are the nesting vectors. In the case of CVHS, there are two sets of nonequivalent nesting vectors $\bm{Q}$ and $\bm{Q}'$.
  }\label{fig:model}
\end{figure*}

Here we explore possible orders in tWSe$_2$ induced by weak repulsive interactions based on renormalization group analysis.
We obtain the phase diagram presented 
in Fig.~\ref{fig:phase}.  
In a large portion of the parameter space
there exist chiral SC states with Chern numbers $\mathcal{N}=1,2,4$, corresponding to $p+ip$, $d+id$ and $g+ig$ states, respectively. For small and large $n$, the ground state is either $f$- or $f'$- wave SC state, corresponding to $A_1$ or $A_2$ representation of the underlying $C_{3v}$ group (or $D_{6h}$ group when $\phi=0$). 
The $i$- and $i'$-wave pairing states with 12 nodes belonging to different representations are also obtained. Intriguingly, at $\phi=\pi/3$ and for small or large $n$, the ground state features pair-density-wave (PDW) superconductivity \cite{doi:10.1146/annurev-conmatphys-031119-050711,larkin1968zh,fulde1964superconductivity,PhysRevLett.88.117001,PhysRevLett.99.127003,cho2012superconductivity,PhysRevX.4.031017,PhysRevB.89.165126,PhysRevLett.114.237001,PhysRevB.76.140505,PhysRevLett.107.187001,PhysRevLett.122.167001,doi:10.1126/sciadv.aat4698,PhysRevB.95.155116,PhysRevLett.113.046402,PhysRevLett.125.167001,Huang2022} with equal-spin pairing, which is degenerate with the opposite-spin pairing at zero momentum when no magnetic field is applied. By applying a weak out-of-plane Zeeman field to the tWSe$_2$, this degeneracy is lifted and the equal-spin pairing (or spin-polarized) PDW is the unique ground state. This spin-polarized PDW state can be probed by experiments including spin-polarized STM measuring spin-resolved pairing gap and electric transport measurement of superconducting diode effect (SDE).
These SC orders survive even approaching the conventional van Hove singularities (CVHSs), shown as the white dashed curve in Fig.\ref{fig:phase}. The special point at $n=1, \phi=\pi/6$ is the higher order van Hove singularity (HOVHS) where the fermion density of states has a power-law divergence. We find the ground state there is a metal without symmetry breaking.


\medskip
{\bf\large\noindent Results}

{\bf Model for \mr superlattice:} The monolayer WSe$_2$ is a triangular lattice semiconductor with broken inversion symmetry \cite{Manzeli2017}. The valence band top is located at $\pm \bm{K}_0$, as shown in Fig. \ref{fig:model}(a). Due to strong spin-orbit coupling, single particle states near $\bm{K}_0$ and $-\bm{K}_0$ have opposite spin polarization. When two layers of WSe$_2$ are AA stacked together and twisted by a small angle, a \mr pattern and a \mr \BZ~develop, as is shown in Fig. \ref{fig:model}(a) \cite{Naik2018,Ruiz-Tijerina2019,Wufc2018,Wufc2019,Naik2020}. The spin up states in the top (down) layer near the $\bm{K}_0$ points are mapped to the states near $\bm{K}$ ($-\bm{K}$) in the \mr \BZ, while the spin down in the top (down) layer near the $-\bm{K}_0$ points are mapped to the states near $-\bm{K}$ ($\bm{K}$).
The hybridization due to the interlayer coupling leads to a narrow \mr band, as shown in Fig. \ref{fig:model}(b). 

Because the twisted bilayer system respects both inversion and time-reversal symmetries, it gives rise to double degeneracy for each band; however, the double degeneracy can be lifted by a finite out-of-plane electric field [see Fig. \ref{fig:model}(c)] through the layer potential difference $V_z$ which explicitly breaks inversion symmetry. A tight binding model for the \mr band can be obtained by constructing a set of Wannier states and fitting with DFT calculations\cite{Wufc2019,Wang2020,Pan2020}. The hopping parameter $t_{ij}$ of electrons with spin polarization $\sigma$ between \mr lattice sites in the presence of a finite out-of-plane electric field picks up a nontrivial phase as $t^\sigma_{ij}=|t_{ij}|e^{i\phi^\sigma_{ij}}$, where $\phi^{-\sigma}_{ij}=-\phi^{\sigma}_{ij}$ required from TRS. The amplitude $|t_{ij}|$ decays exponentially with distance between $i$ and $j$, which allows for a nearest-neighbor hopping approximation; namely $|t_{ij}|=t$ on NN bonds and zero otherwise. The spin-dependent phase on NN bonds is $\pm \phi$, giving rise to an accumulated flux $\pm 3\phi$ on each triangular plaquette, as shown in Fig. \ref{fig:model}(d). The magnitudes of $\phi$ depends on the strength of electric field (or $V_z$) in a monotonic way \cite{Pan2020}, and was shown in Fig. \ref{fig:model}(e) for clarity. 

Moreover, it is found the onsite Coulomb repulsion is much larger than the interaction between two adjacent sites \cite{Pan2020,Wang2020}, which validates the following triangular-lattice Hubbard model description:
\begin{equation}
  H=\sum_{\bm{k},\sigma=\pm}\epsilon_{\bm{k}}^\sigma c_{\bm{k}\sigma}^\dagger c_{\bm{k}\sigma}+U\sum_{i}n_{i\up}n_{i\down},\label{eq:H}
\end{equation}
where the single-particle dispersion is given by 
$\epsilon_{\bm{k}}^\sigma=-2t\sum_{\bm{a}_m}\cos(\bm{k}\cdot\bm{a}_m+\sigma\phi)$. 
Hereafter we set $\bm{a}_1=(1,0)$, $\bm{a}_2=(-1/2,\sqrt{3}/2)$, $\bm{a}_3=(-1/2,-\sqrt{3}/2)$ in unit of \mr lattice constant.
For generic $\phi\neq 0$, the model respects time-reversal symmetry, spin-rotational symmetry along the z axis, and lattice point-group symmetry $C_{3v}$ (for $\phi=0$ the point-group symmetry is $D_{6h}$). Moreover, the model has the following symmetries regarding $\phi$. 
First, the model is invariant by interchanging spin polarizations and changing $\phi$ to $-\phi$. In addition, changing the total flux on each triangular plaquette by $2\pi$ should leave the phase diagram invariant, and this corresponds to shifting $\phi$ by $\pm2\pi/3$. Finally, particle-hole transformation leads to $n\to 2-n$ and $\phi \to \pi-\phi$. 
Consequently, the ground state phase diagram of the model is symmetric under $\phi\to\phi\pm\pi/3$ and $n\to2-n$\cite{Zhang2021}.

{\bf Competing orders.} For each $\phi$, there is a doping with VHS on the FS \cite{Zhang2021}. When $\phi\neq \pi/6$, there are three nonequivalent CVHSs for each spin, as shown in Fig.\ref{fig:model}(h), which merge into a single HOVHS at $\phi=\pi/6$ [Fig.\ref{fig:model}(i)], where the DOS has a power-law divergence. The perfectly nested FS and divergent DOS promote the density-waves (DW) as competing orders as opposed to the SC order. Due to the divergent DOS at these VHSs, it suffices look into the small patches around each VHS. We thereby employ the parquet renormalization group (pRG)\cite{Zheleznyak1997,Furukawa1998,Chubukov2008,Nandkishore2012,Lin2019,Hsu2021} analysis based on the six patch model for the CVHS and the two patch model for the HOVHS to study the leading instabilities. Technical details can be found in Supplementary Information.

\begin{figure}
  \includegraphics[width=7.5cm]{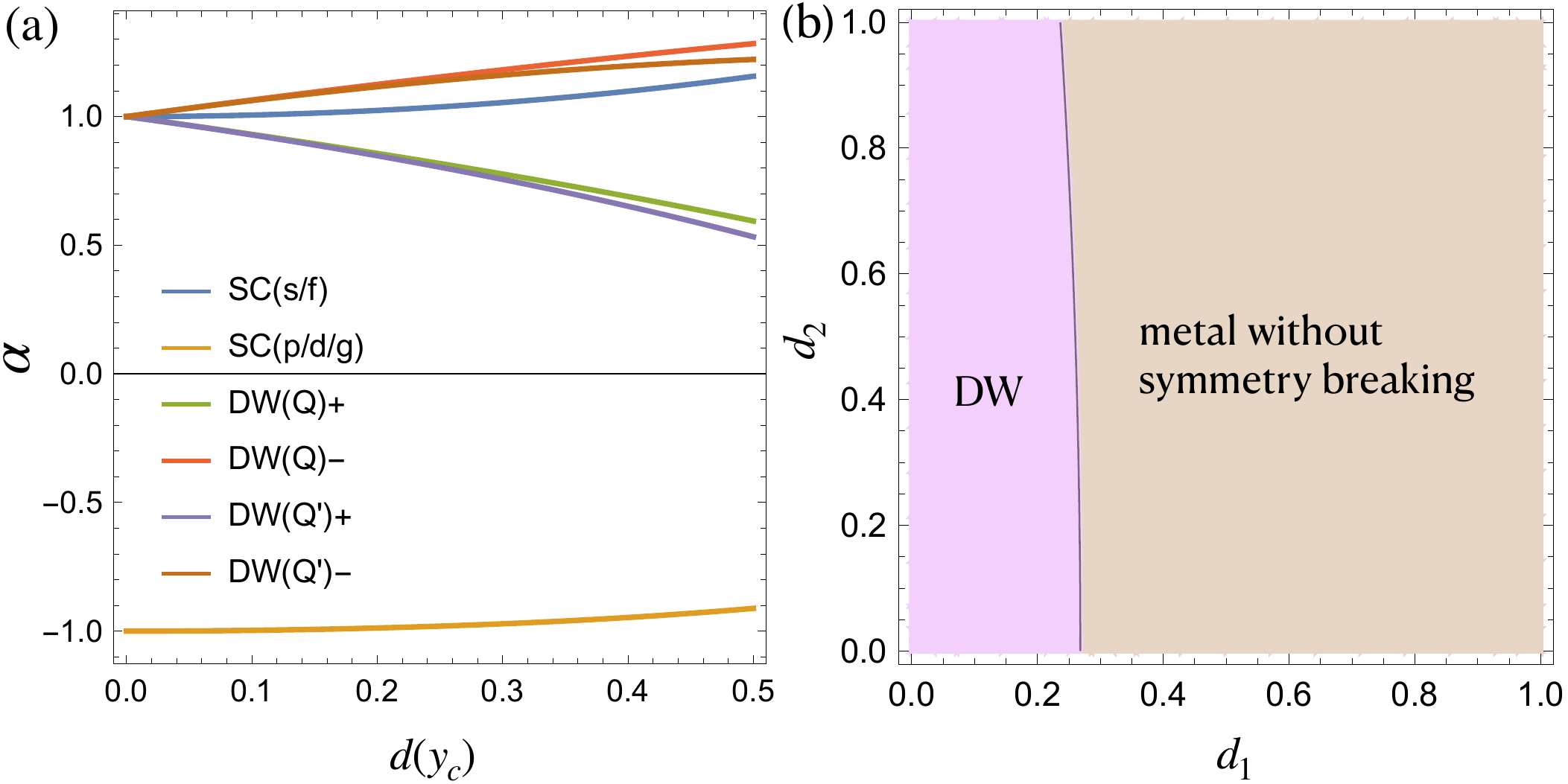}
  \caption{(a)RG results for the case of CVHS near perfect nesting. The susceptibility for each order has a scaling form $\chi\sim(y-y_c)^\alpha$. Therefore SC(p/d/g) is the only possible order in low $T$ limit ($\pm$ in DW$\pm$ means even and odd parity). (b) Two-loop RG results of HOVHS at $\phi=\pi/6$. There is no symmetry breaking near perfect nesting, which is characterized by an interacting RG fixed point and dubbed as 'Supermetal' in \cite{Isobe2019}. DW wins over SC at small nesting beyond a critical interaction, but SC is the only instability in the weak coupling limit when doped away from van Hove filling.}\label{fig:co}
\end{figure}

For the six patch model, both the bare particle-hole (ph) susceptibilities $\Pi_{ph}(\bm{Q})$ and $\Pi_{ph}(\bm{Q}')$ and the particle-particle (pp) susceptibility $\Pi_{pp}(0)$ scale as $\log^2(\Lambda/T)$ where $\Lambda$ is the ultraviolet cutoff. To develop the pRG equations, we use $y=\Pi_{pp}(0)$ as the running parameter, and define $d_{1}=d\Pi_{ph}(\bm{Q})/dy$ and $d_{2}=d\Pi_{ph}(\bm{Q}')/dy$ as the nesting parameters. For hexagonal lattices the maximum value of $d$ at perfect nesting is $1/2$\cite{Lin2019}. 
Taking the Hubbard interaction $U\approx1t$ as the initial input, we find there is a critical value $y_c$ where all the interactions flows to strong coupling limit. Near this critical point, 
the susceptibilities for various order parameters have the scaling form $\chi\sim(y-y_c)^\alpha$, which signals the onset of some order only if $\alpha<0$, and the most negative $\alpha$ corresponds to the leading order. In Fig.\ref{fig:co}(a) we plot $\alpha$ for different competing orders as a function of $d(y_c)=d_1(y_c)\approx d_2(y_c)$. In all range of $d(y_c)$, the $p/d/g$-wave SC, which belongs to $E$ representation, is the only possible order in low $T$ limit. Our result is consistent with the $\phi\to0$ limit shown in Ref.\cite{Nandkishore2012}.

For the two patch model of HOVHS, all $\Pi_{pp}(0)$, $\Pi_{ph}(0)$ and $\Pi_{ph}(\bm{Q})$ scale as $1/T^{1/3}$. We again use $y=\Pi_{pp}(0)$ as the running parameter for the pRG analysis and define $d_1=\Pi_{ph}(\bm{Q})/y$ and $d_2\approx3\Pi_{ph}(0)/y$ as the nesting parameters ($d_1=d_2=1$ for perfect nesting). The only interaction involved in this case is the inter-valley density interaction. At perfect nesting, the one-loop pRG equation vanishes, indicating the importance of higher order contributions. We thus calculate the pRG equations up to two-loop level. Interestingly, for repulsive interaction the system does not flow to strong coupling limit, implying no symmetry breaking. In fact, there is an interacting fixed point with $d_1=1$ at Van Hove doping, which is a non-Fermi liquid and was dubbed as `supermetal' in Ref\cite{Isobe2019}.
The system becomes Fermi liquid when doped away from Van Hove filling but still in the gapless regime. The phase diagram is shown in Fig.\ref{fig:co}(b).  
Away from perfect nesting, when $d_1\lesssim0.25$, the DW order wins over SC beyond a critical interaction. But the SC instability is the only instability in the weak coupling limit in this regime.

Below we focus on the SC order, and employ the Raghu-Kivelson-Scalapino RG analysis \cite{Raghu2010,Cho2013} to identify the leading pairing channel for the ground state. The results are summarized in Fig.\ref{fig:phase}. In the following, we discuss two particularly interesting cases: pair-density-wave and chiral superconductivity, respectively, and leave the technical details in Supplementary Information.

\begin{figure}[t]
  \includegraphics[width=8.5cm]{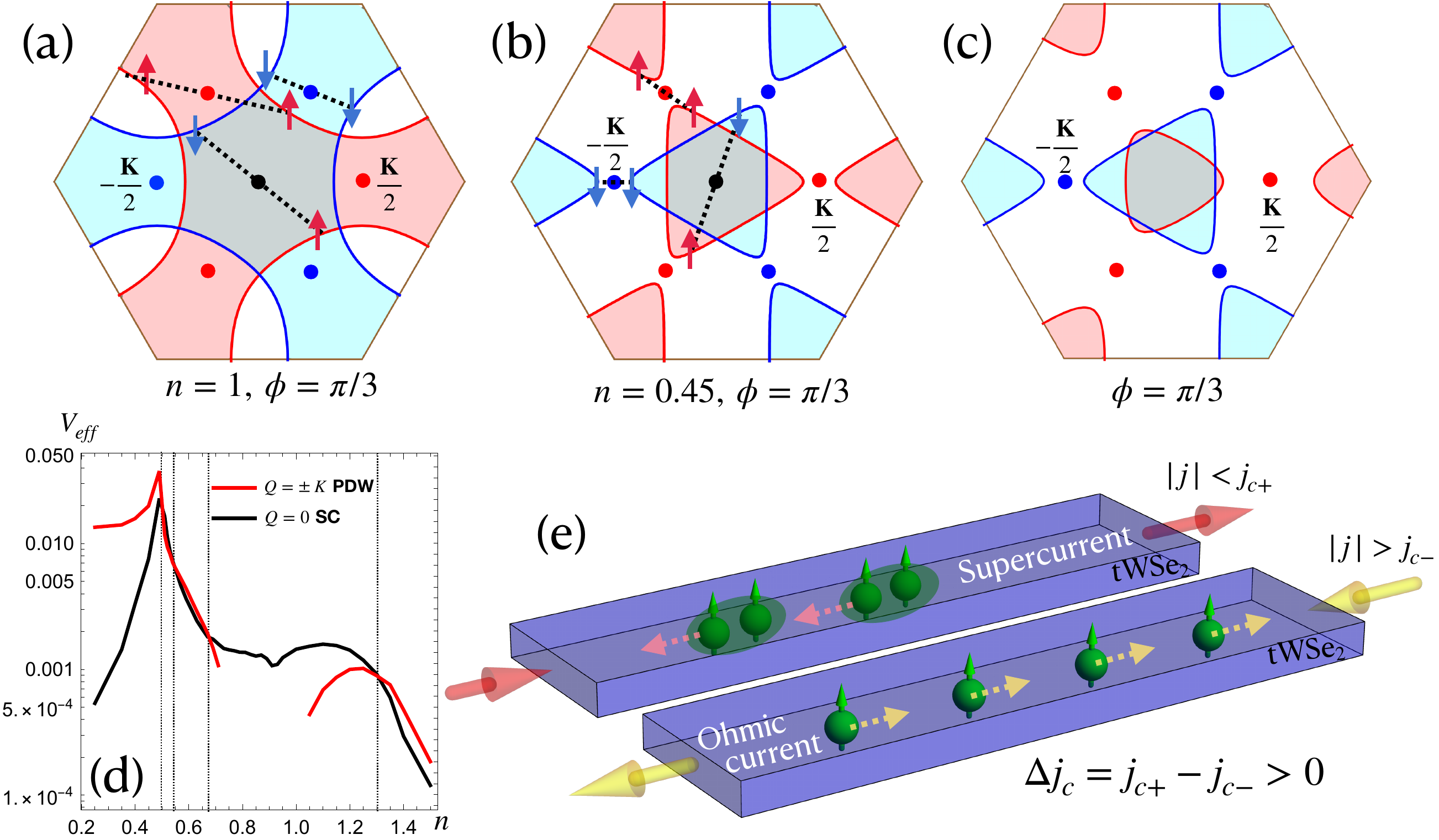}
  \caption{(a) and (b). Emergence of finite $\bm{Q}$ pairing at $\phi=\pi/3$.  (c) Fermi surface in the presence of an out-of-plane Zeeman field at $\phi=\pi/3$.  (d) Pairing strength of $\bm{Q}=\pm\bm{K}$ PDW order and $\bm{Q}=0$ SC state at $\phi=\pi/3$.(e) Experimental setup for measuring the spin polarized PDW using SDE.}\label{fig:pdw}
\end{figure}

{\bf Pair-density-wave:} For the particular case $\phi=\pi/3$ (more generally $\pm\pi/3$ modulo $\pi$), both spin-up and spin-down FSs are nested in the pp channel; namely 
FSs for each spin are symmetric with respect to the finite momentum points at $\pm \bm K/2$ [shown as blue and red points in Fig.~\ref{fig:pdw}(a) and (b)], rending to possible pairing whose center of mass momentum is finite with $\bm{Q}=\pm \bm{K}$, i.e. a PDW superconductivity. Indeed, for $0<n<0.5$, $0.54<n<0.68$ and $1.29<n<2$, our RG analysis shows that the ground state features same-spin pairing at finite momentum $\pm \bm{K}$ and opposite-spin pairing at zero momentum, which are degenerate in energy. 

The emergence of PDW order of pairing between electrons with the same spin polarization at $\phi=\pi/3$ is actually 
no 
surprise because this particular $\phi=\pi/3$ is related to $\phi=0$ by symmetry transformation: a particle-hole transformation plus a local gauge transformation which changes $\phi\to \phi\pm \pi/3$ and $n\to 2-n$. 
Under this transformation a triplet pairing with total $S^z=\pm 1$ at $\phi=0$ and $2-n$ is mapped to a PDW order at $\pi/3$ and $n$. To see how this happens, consider the model with $\phi=0$ featuring a zero-momentum triplet pairing ground state, 
for which the triplet order with $S^z=+1$ 
have the form of $\Delta_{\uparrow\uparrow}(\bm{R},\bm{r})c_{i\uparrow}c_{j\uparrow}$ [$\bm{R}=(\bm{r}_i+\bm{r}_j)/2$ the position of the center of mass, and $\bm{r}=\bm{r}_i-\bm{r}_j$ the relative position]. In momentum space, this fermion bilinear term is $\Delta_{\uparrow\uparrow}(\bm{Q},\bm{k})c_{\uparrow}(\bm{k}-\bm{Q}/2)c_{\up}(-\bm{k}-\bm{Q}/2)$. For $\phi=0$, triplet pairing has zero momentum $\bm{Q}=0$ and its real space order parameter is independent of $\bm{R}$. After the particle-hole transformation ($c_{j\sigma}\to c^\dag_{j\sigma}$) and a local gauge transformation ($c_{j\sigma}\to e^{i\sigma\eta_j}c_{j\sigma}$ with $\eta_i=\bm{K}\cdot \bm{r}_i$), 
the corresponding order parameter acquires a spatial phase modulation,
namely,
\begin{equation}
  \Delta_{\up\up}(\bm{r})\to\Delta_{\up\up}(\bm{r})e^{i(\eta_i+\eta_j)}=\Delta_{\up\up}(\bm{r})e^{i\bm{Q}\cdot\bm{R}}
\end{equation}
where $\bm{Q}=\bm{K}=(\frac{4\pi}{3},0)$. 
For pairing between two spin down fermions, the argument is completely parallel, but we will have $\bm{Q}=-\bm{K}$ instead. One can further see that for the opposite-spin pairing ($S^z=0$) of the model with $\phi=0$ is mapped to usual zero-momentum paring of the model with $\phi=\pi/3$. 
This is because spin up and spin down sectors transform in opposite direction, 
  $\Delta_{\up\down}(\bm{r})\to\Delta_{\up\down}(\bm{r})e^{i(\eta_i-\eta_j)}\sim\Delta_{\up\down}(\bm{r})$,
which does not lead to non-trivial $\bm{R}$ dependence. 
Therefore, an opposite-spin pairing state ($S^z=0$) at $\phi=0$ is still a $\bm{Q}=0$ SC state at $\phi=\pi/3$, while a triplet pairing state with $S^z=\pm 1$ at $\phi=0$ transforms into a $S_z=\pm 1$ PDW order at $\phi=\pi/3$. 

Since at $\phi=0$ the same-spin triplet pairing with $S^z=\pm 1$ is degenerate with the opposite-spin triplet pairing with $S^z=0$ due to the full spin $SU(2)$ rotational symmetry, the system at $\phi=\pi/3$ also maintains this degeneracy between the same-spin PDW order and the opposite-spin zero-momentum pairing.
The degeneracy between finite-$\bm{Q}$ PDW and zero-$\bm{Q}$ SC state can be lifted by applying a weak out-of-plane magnetic field, which generates a Zeeman coupling to the spin of electrons. 
Moreover, for the twisted angle $\theta\gtrsim3^{\circ}$, the orbital effect of the applied magnetic field can be neglected \cite{Zhang2021}, leaving the Zeeman coupling the only dominant effect. 
In Fig.~\ref{fig:pdw}(c) we draw as an example the FS configurations with a finite Zeeman coupling, which differentiates the sizes of the spin up and spin down FSs. 
As a result, the $\bm{Q}=0$ pairing between $\bm{k}$ and $-\bm{k}$ is suppressed, and the PDW becomes the unique ground state. 
Moreover, the degeneracy between spin-up and spin-down PDW is also lifted by the Zeeman field: fermions with the larger FS tend to have a stronger pairing strength and hence a higher transition temperature. 
To show the competition between the spin-polarized PDW and another $\bm{Q}=0$ SC state, in Fig.~\ref{fig:pdw}(d) we plot the dimensionless pairing strength $V_{\text{eff}}$ for the PDW order as well as the $\bm{Q}=0$ SC state in a weak magnetic field. $V_{\text{eff}}$ is defined such that $T_c\approx W\exp[-1/(V_{\text{eff}}U^2/t^2)]$.For an estimation we take $t=5meV\approx58K$, and $W=9t$, $U=2.5t$, then $T_c$ ranges from $0.18K$ to $9.56K$ when $V_{\text{eff}}\in(0.02,0.04)$. From the result we see $V_{\text{eff}}$ at small $n$ is much larger than that at large $n$. 
The PDW order has been suggested as intrapocket pairing in monolayer system \cite{Hsu2017}.

Here in our case, the PDW order for large $n$ can also be interpreted as intrapocket pairing, while the at small $n$, it is the interpocket pairing between electrons with the same spin polarizations that leads to PDW. 
The pairing strength of interpocket PDW for small $n$ has higher $T_c$ and is more promising to realize. 

This PDW state has the superconducting diode effect (SDE) due to the absence of inversion and time reversal symmetry [See \ref{fig:pdw}(e) for clarity], similar to that of the Flude-Fellel state \cite{PhysRevLett.128.037001}. The critical currents parallel to the direction of $\bm{Q}$, is different from that in the opposite direction. As a result, an external depairing current can induce Ohmic current in one direction but remains supercurrent in the opposite direction.

\begin{figure}
  \includegraphics[width=6.5cm]{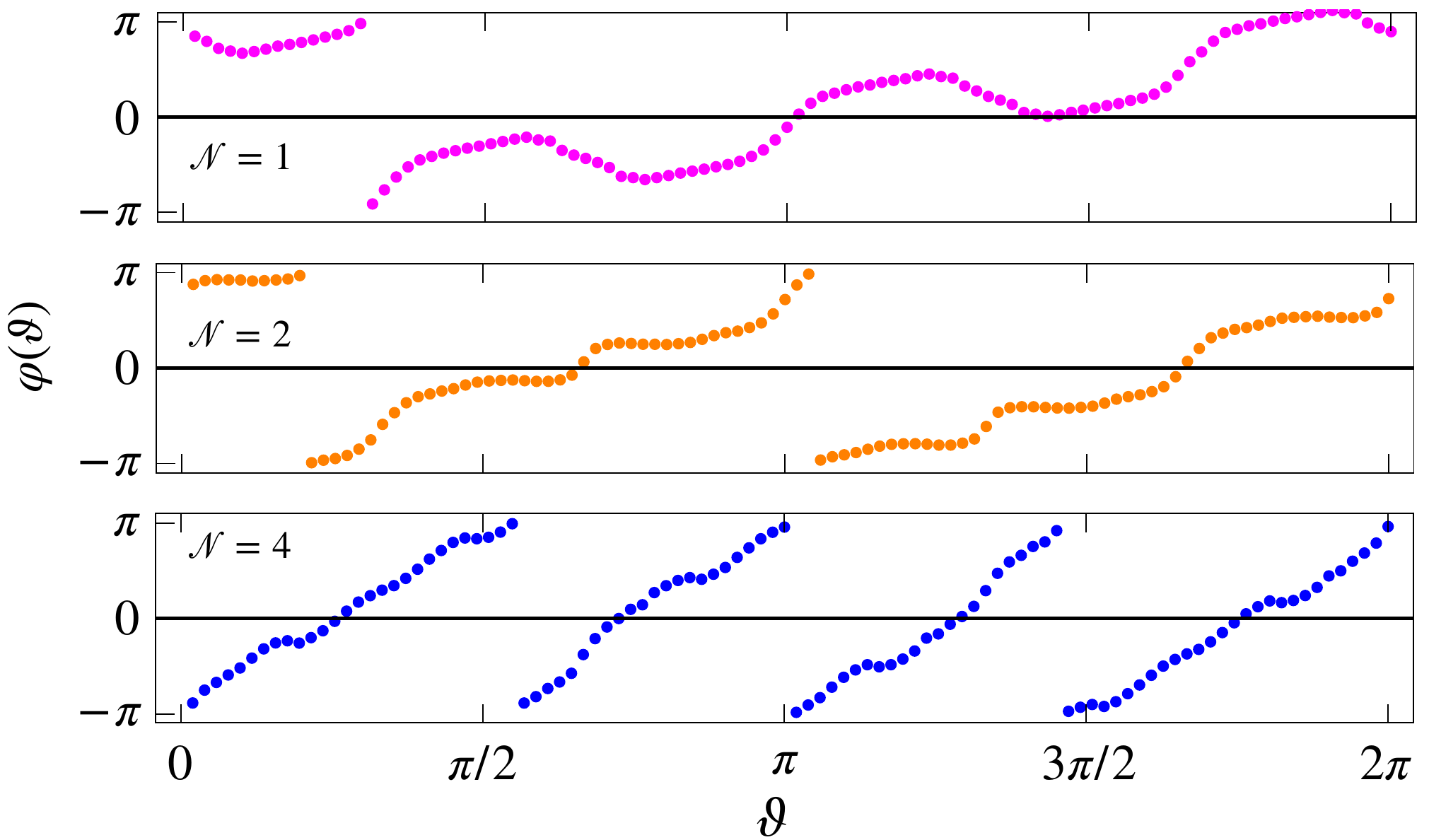}
  \caption{SC phase variation along the FS parametrized by the angle $\vartheta$. Distinct topological phases are characterized by the number of $2\pi$ slips when moving around the FS in a full circle. }\label{fig:chiral}
\end{figure}

{\bf Chiral superconductivity:} In a large part of the parameter space 
around $n=1$, we find two degenerate pairing states with mixed parity, which form the basis of  
the two dimensional $E$ representation.
Thus, a general state can be written as
$\Delta(\bm{k})=\Delta_1v_1(\bm{k})+\Delta_2v_2(\bm{k})$, 
where $v_1$ and $v_2$ are two {\it real} orthonormal basis satisfying $\text{Tr}v_1^2=\text{Tr}v_2^2=1,\text{Tr}(v_1v_2)=0 $, and the trace here is the shorthand for momentum integration.

The true ground state should minimize the Ginzburg-Landau free energy, and this helps to determine the two complex amplitudes $\Delta_1$ and $\Delta_2$. 
To this end, we start with the action in pairing channel from \eqref{eq:H}, and then perform Hubbard-Stratonavich transformation by introducing the gap function $\Delta(\bm{k})$ as an auxiliary field. 
Integrating out fermions and expanding $\Delta$ to the quartic order, we obtain,
\begin{equation}
  \begin{aligned}
    \mathcal{F}[\Delta_1,\Delta_2]&=\alpha(T-T_c)(|\Delta_1|^2+|\Delta_2|^2)\\
    &+\beta_1(|\Delta_1|^2+|\Delta_2|^2)^2+\beta_2|\Delta_1^2+\Delta_2^2|^2\label{eq:F2}
  \end{aligned}
\end{equation}
where $\alpha$ is a positive constant, 
$\beta_1=(1/3)K\tr(v_1^4)$ and $\beta_2=(1/2)K\tr(v_1^2v_2^2)$ with $K>0$ (See Supplementary Information).
Since $v_1$ and $v_2$ are real 
, we always have $\beta_1>0$ and $\beta_2>0$. 
This fact leads to nontrivial consequences. 
First $\beta_1$ being positive guarantees the existence of a SC ground state. 
Furthermore, in order to minimize the free energy with $\beta_2>0$, the term $|\Delta_1^2+\Delta_2^2|^2$ must vanish. 
It vanishes only when $|\Delta_1|=|\Delta_2|$ and $\arg(\Delta_1/\Delta_2)=\pi/2$, corresponding to a chiral SC state: the phase of pairing $\varphi=\arg[\Delta(\bm{k})]$ winds up multiple times of $2\pi$ when $\bm{k}$ goes around the whole FS while the amplitude $|\Delta(\bm{k})|$ remains nonzero on the FS (nodeless), and it spontaneously breaks the time reversal symmetry.

The nodeless chiral superconductor is topological, which supports chiral fermionic modes on edges of the system. 
The topological invariant is characterized by the Chern number defined as: 
\begin{equation}
  \mathcal{N}=\frac{1}{4\pi}\int_{BZ}d\bm{k}[ \hat{\bm{h}}\cdot(\partial_{k_x}\hat{\bm{h}}\times\partial_{k_y}\hat{\bm{h}})]\label{eq:chern},
\end{equation}
where $\hat{\bm{h}}=(\Re[\Delta(\bm{k})],\Im[\Delta(\bm{k})],\xi_{\bm{k}})/E_{\bm{k}}$ and $E_{\bm{k}}=\sqrt{\xi_{\bm{k}}^2+\Delta^2(\bm{k})}$. 
$2\pi\mathcal{N}$ is the Berry flux of the two-level Hamiltonian $h_k=\hat{h}(\bm{k})\cdot\bm{\sigma}$, or the monopole charge located at the torus center, which is just the line integral of a gauge transformation on the FS, according to the Stokes theorem. In other words, Eq.\eqref{eq:chern} is the same as the winding numbers defined by the number of times that the SC phase $\varphi$ slips $2\pi$ when $\bm{k}$ sweeps around the FS \cite{Black_Schaffer_2014}. 
There will be $2\mathcal{N}$ chiral Majorana edge modes or $\mathcal{N}$ chiral complex fermion edge modes \cite{Sato_2017,PhysRevB.61.10267,PhysRevLett.109.197001,Black_Schaffer_2014}.
In Fig. \ref{fig:chiral} we show some examples of the chiral SC states obtained in our model. We plot the SC phase $\varphi$ as a function of the FS parameter $\vartheta$. Since the FS here forms a closed loop (either centered at $\bm{\Gamma}$ point or $\pm\bm{K}$ point), we can parametrize the points on FS by the angle $\vartheta$ formed by $\bm{k}_F$ (or $\bm{k}_F\pm\bm{K}$) and $\hat{x}$.

{\bf Discussions and concluding remarks:} In this paper, we have shown that an out-of-plane electric field and magnetic field induced spin-polarized PDW order can arise in the twisted bilayer TMD system, which is the unique ground state. This PDW state supports nonzero spin-polarized SC diode current, which can be directly probed using the SDE experiment \cite{Ando2020,Lyu2021,Baumgartner2022}.  
We also find various topological chiral SC states in this system with the Chern number $|\mathcal{N}|=1,2,4$. 
The model we studied is experimentally accessible and is promising to be realized in experimental setups.
Various other exciting physics, such as the charge-4e superconductor or possible quantum critical behavior between different phases tuned by the out-of-plane electric field, is left for future study.

\medskip
{\bf\large\noindent Data Availability.} 

The authors declare that all data supporting the findings of this study are available within the paper and its supplementary information file.

\medskip
{\bf\large\noindent References.}
\bibliography{twisted}

\medskip
{\bf\large\noindent Acknowledgement}

We would like to sincerely thank Linhao Li, Zi-Xiang Li, Fengcheng Wu, and Ya-Hui Zhang for useful discussions. This work is supported in part by the NSFC under Grant No. 11825404 (YMW, ZZW and HY), the MOSTC Grant No. 2018YFA0305604 (HY), the CAS Strategic Priority Research Program under Grant No. XDB28000000 (HY). YMW is also supported in part by Shuimu Fellow Foundation at Tsinghua.

\medskip
{\bf\large\noindent Author Contributions.} 

H.Y. designed and supervised this work. Y.-M.W. and Z.W. carried out the RG analysis analytically and numerically. All authors contribute to writing this paper.

\medskip
{\bf\large\noindent Competing Interests.}

All authors declare no competing interests in this work.

\onecolumngrid
\vspace{1cm}
\begin{center}
{\bf\large Supplemental Materials}
\end{center}

\setcounter{equation}{0}
\setcounter{figure}{0}
\setcounter{table}{0}
\makeatletter
\renewcommand{\theequation}{S\arabic{equation}}
\renewcommand{\thefigure}{S\arabic{figure}}
\renewcommand{\bibnumfmt}[1]{[S#1]}
\renewcommand{\citenumfont}[1]{S#1}

\subsection{A. Competing orders at the VHS line} 

\label{sub:competing_orders_at_the_vhs_line}
\begin{figure}
  \includegraphics[width=16cm]{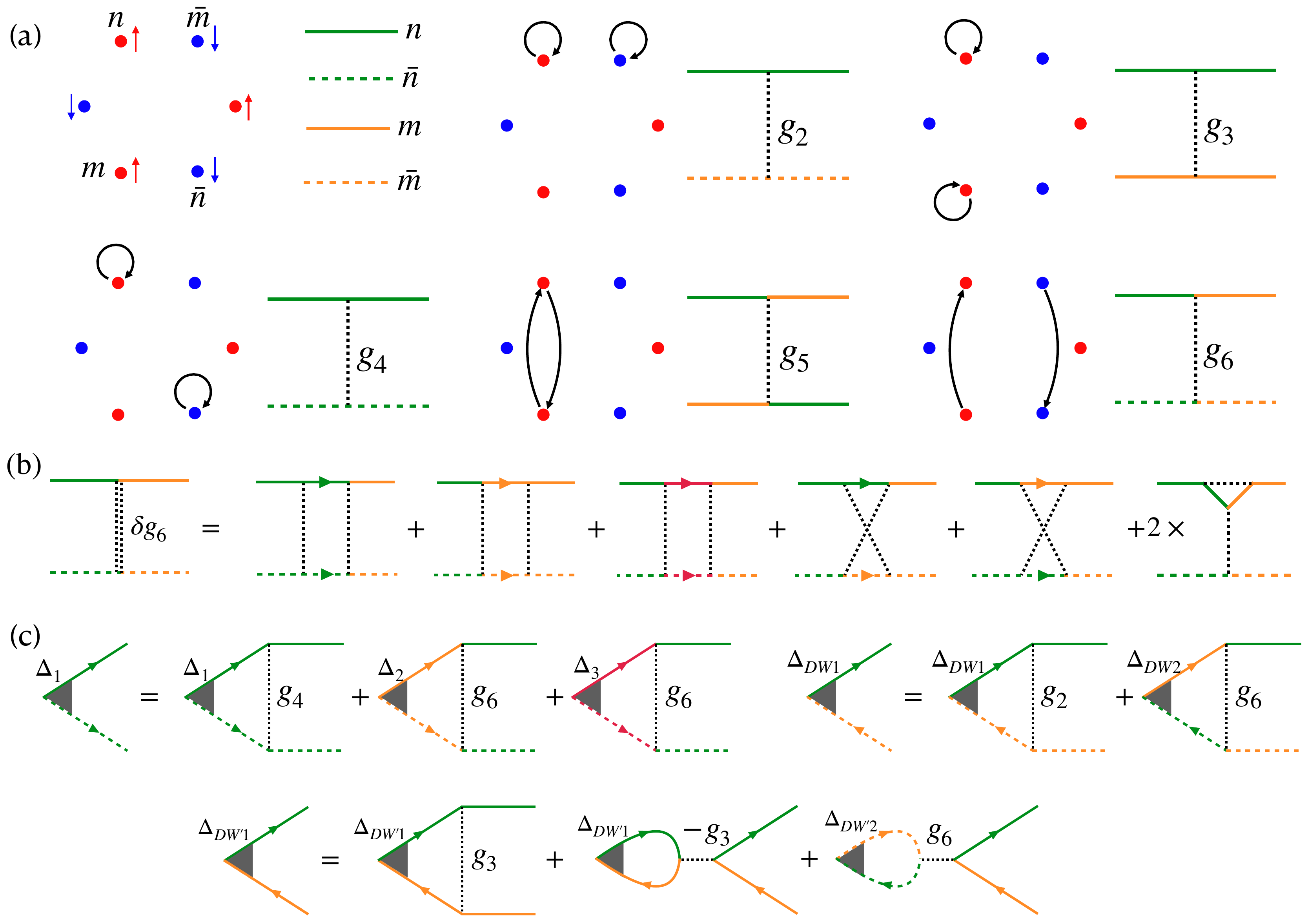}
  \caption{(a) Illustration of the six patch model for the case of conventional VHS. The patch index $n$ and $m$ are for spin up fermions and $\bar{n}$ and $\bar{m}$ for spin down fermions. Due to momentum conservation and spin $S_z$ preservation, we consider five different interactions. (b) Diagrammatic representation of the RG equations for $g_6$. (c) Diagrammatic representation of the RG equations for the SC order parameter, i.e. \eqref{eq:beta_SC}, and the DW order parameter peaked at $\bm{Q}$ and $\bm{Q}'$, i.e. \eqref{eq:beta_DW} and \eqref{eq:beta_DWp}.} \label{fig:VHS}
\end{figure}

When the system is tuned close to the van Hove singularity line, as shown in Fig.\ref{fig:VHS}(a), the Fermi surfaces are nested. Depending on the value of $\phi$ (and hence the electric field), there can be six or two different VHSs. The case with two different VHSs occurs at $\phi=\pi/6$, and these VHS are higher order ones, which give rise to power-law divergence in the density of states, instead of the $\log$ divergence in the conventional VHS case. The Fermi surface nesting enhances the particle-hole channel susceptibilities, which can lead to possible spin or charge density waves and compete with the Cooper channel superconductivity. Thus, we need to treat the superconducting order and density wave order on equal footing. An unbiased way for achieving this is the parquet RG (pRG). Since the density of states at the VHS points are much larger than that in the rest of the \BZ, we can only consider small patches around these VHS points. Therefore, we will be dealing with a six-patch model for $\phi\neq\pi/6$ and a two-patch model for $\phi=\pi/6$.

In the six patch model when $0<\phi<\pi/6$ (for $\pi/6<\phi<\pi/3$ similar six patch analysis yields the same results), the three van Hove points [see Fig.\ref{fig:VHS}(a)] for the spin up FS are located at
\begin{equation}
  \bm{1}=(2\phi-\pi,2\sqrt{3}\phi+\frac{\pi}{\sqrt{3}}), ~\bm{2}=(-4\phi,\frac{-2\pi}{\sqrt{3}}),~\bm{3}=(\pi+2\phi,\frac{\pi}{\sqrt{3}}-2\sqrt{3}\phi).
\end{equation}
  The dispersions near these points are:
\begin{equation}
  \epsilon_1=\tilde{t}(-k_x^2+\sqrt{3}k_xk_y),~ \epsilon_2=\tilde{t}(\frac{1}{2}k_x^2-\frac{3}{2}k_y^2),~\epsilon_3=\tilde{t}(-k_x^2-\sqrt{3}k_xk_y)
\end{equation}
where $\tilde{t}=t\cos(3\phi)$. We see $\tilde{t}$ vanishes at $\phi\to\pi/6$, we thus need to keep higher order terms in the expansion. Similar expansions for spin down sector near $\bar{\bm{1}}=-\bm{1}$, $\bar{\bm{2}}=-\bm{2}$ and $\bar{\bm{3}}=-\bm{3}$ show that $\epsilon_{\bar{1}}=\epsilon_1$, $\epsilon_{\bar{2}}=\epsilon_2$ and $\epsilon_{\bar{3}}=\epsilon_3$.
The building block for the RG calculation is the $\log^2$ divergence of both the particle-particle and particle-hole channel susceptibilities,
\begin{equation}
   \begin{aligned}
     &\Pi_{pp}(0)\approx h^{pp0}\log\frac{\Lambda}{\max(T,|\mu|)}\log\frac{\Lambda}{T}, ~\Pi_{ph}(\bm{Q})\approx\Pi_{ph}(\bm{Q}')\approx h^{phQ}\log\frac{\Lambda}{\max(T,|\mu|)}\log\frac{\Lambda}{\max(T,|\mu|,t')}\\
   \end{aligned}
 \end{equation} 
 where $\Lambda$ is the ultraviolet cutoff and the chemical potential $\mu$ measures the distance from perfect nesting: $\mu=0$ corresponds to the doping right at the VHS. $t'$ denotes the long distance hopping terms that spoil the nesting. For square lattice, the two factors $h^{pp0}$ and $h^{phQ}$ are identical. For the hexagonal lattices such as the triangular lattice that we consider here, it is easy to see,
 \begin{equation}
    h^{hpQ}=\frac{1}{2}h^{pp0}=\frac{1}{4\sqrt{3}\pi^2t\cos(3\phi)}.
\end{equation} 
Thus the maximum value of the ratio $\Pi_{ph}(\bm{Q})/\Pi_{pp}(0)$ and $\Pi_{ph}(\bm{Q}')/\Pi_{pp}(0)$ is $1/2$, achieved at $\mu=0$ and $t'=0$, i.e. at the perfect nesting point. Note that at the perfect nesting, the two p-h bubbles $\Pi_{ph}(\bm{Q})$ and $\Pi_{ph}(\bm{Q}')$ are identical.

In the two patch model when $\phi=\pi/6$, the quadratic terms in the energy expansion near the two VHS points vanish, leading to a higher order VHS where the energy dispersions are 
\begin{equation}
  \epsilon_1=\frac{t}{4}k_x(k_x^2-3k_y^2), ~\epsilon_2=-\frac{t}{4}k_x(k_x^2-3k_y^2).
\end{equation}
Unlike the six patch case, here because of the power law divergence in the density of states, we also need to consider the Pomeranchuk instability in the particle-hole channel, as well as the finite momentum pairing in the particle-particle channel. Therefore, we will utilize the following four susceptibilities as our RG building blocks:
\begin{equation}
  \Pi_{pp}(0)=\frac{v^{pp0}}{T^{1/3}},~\Pi_{pp}(\bm{Q})=\frac{v^{ppQ}}{T^{1/3}},~\Pi_{ph}(0)=\frac{v^{ph0}}{T^{1/3}},~ \Pi_{ph}(\bm{Q})=\frac{v^{phQ}}{T^{1/3}},~
\end{equation}
where we have
\begin{equation}
  3v^{ppQ}=3v^{ph0}\approx v^{pp0}=v^{phQ}=3.56/t^{2/3}
\end{equation}
Having identified the divergent susceptibilities, below we show the pRG analysis for the two models separately.

\subsubsection{Six patch model} 
For $\phi>0$ and $\phi\neq\pi/6$, the system doped to the VHS can be described by a six-patch model depicted in Fig.\ref{fig:VHS}(b). The conservation of both momentum and the $z$-component of the spin allows for five different kinds of interactions, namely
\begin{equation}
 \begin{aligned}
    H_I&=\frac{1}{2}g_4\sum_{n\in\{i\}}\rho_n\rho_{\bar{n}}\\
    &+\frac{1}{2}\sum_{m\neq n; {m,n}\in \{i\}}\left(g_2\rho_n\rho_{\bar{m}}+g_3\rho_n\rho_{m}+g_5\psi^\dagger_n\psi^\dagger_m\psi_n\psi_m+g_6\psi^\dagger_n\psi^\dagger_{\bar{n}}\psi_{\bar{m}}\psi_m\right).
 \end{aligned}
\end{equation}
Here $\rho_n=\psi_n^\dagger\psi_n$ is the density operator and $i=1,2,3$ denotes one of the three patches for, e.g. spin up FS, shown in the first plot of Fig.\ref{fig:VHS}(a). Note we are dealing with an effectively spinless model, thus we neglect the intra-patch density-density interaction $g_1$. This term becomes important in spinful model. Moreover, for spinless model, $g_3$ and $g_5$ are in fact the same interaction but in opposite sign, i.e. $g_3=-g_5$. For clarity we also show the Feynman diagrams for all these interactions in Fig.\ref{fig:VHS}(a), where we have assigned a solid line for the spin-up fermions and a dashed line for the spin-down fermions. The different color then are used to distinguish different patches. Under this convention the conservation of $S_z$ can be clearly seen from these diagrams: a solid line cannot be scattered into a dashed line and vice versa. 

The one-loop RG equations for these interactions can be easily obtained. Given that the leading divergence is the double-$\log$ contributions from $\Pi_{ph}(\bm{Q})$, $\Pi_{ph}(\bm{Q}')$ and $\Pi_{pp}(0)$, we only need to keep the one-loop diagrams that contain particle-hole bubbles with momentum transfer $\bm{Q}$ and $\bm{Q}'$ and the particle-particle bubbles with zero momentum transfer. Diagrammatically, the particle-hole bubbles that contribute must have a solid line on one side and a dashed line on the other, and these two lines must have different colors; or they must be both solid or both dashed, but again with different colors. The particle-particle bubbles that contribute must be single colored, and must have both solid and dashed lines. Applying these rules, it's straightforward to obtain
\begin{equation}
  \begin{aligned}
    \dot{g_2}&=d_1(g_2^2+g_6^2)\\
     \dot{g_3}&=d_2(g_3^2+g_6^2)\\
    \dot{g_4}&=-g_4^2-2g_6^2\\
    \dot{g_6}&=-2g_4g_6-g_6^2+2d_1g_2g_6+2d_2g_3g_6\\
  \end{aligned}\label{eq:betaf}
\end{equation}
As an example, we show the equation for $g_6$ diagrammatically in Fig.\ref{fig:VHS}(b). Other equations can be obtained in a similar manner.
\begin{figure}
  \includegraphics[width=14cm]{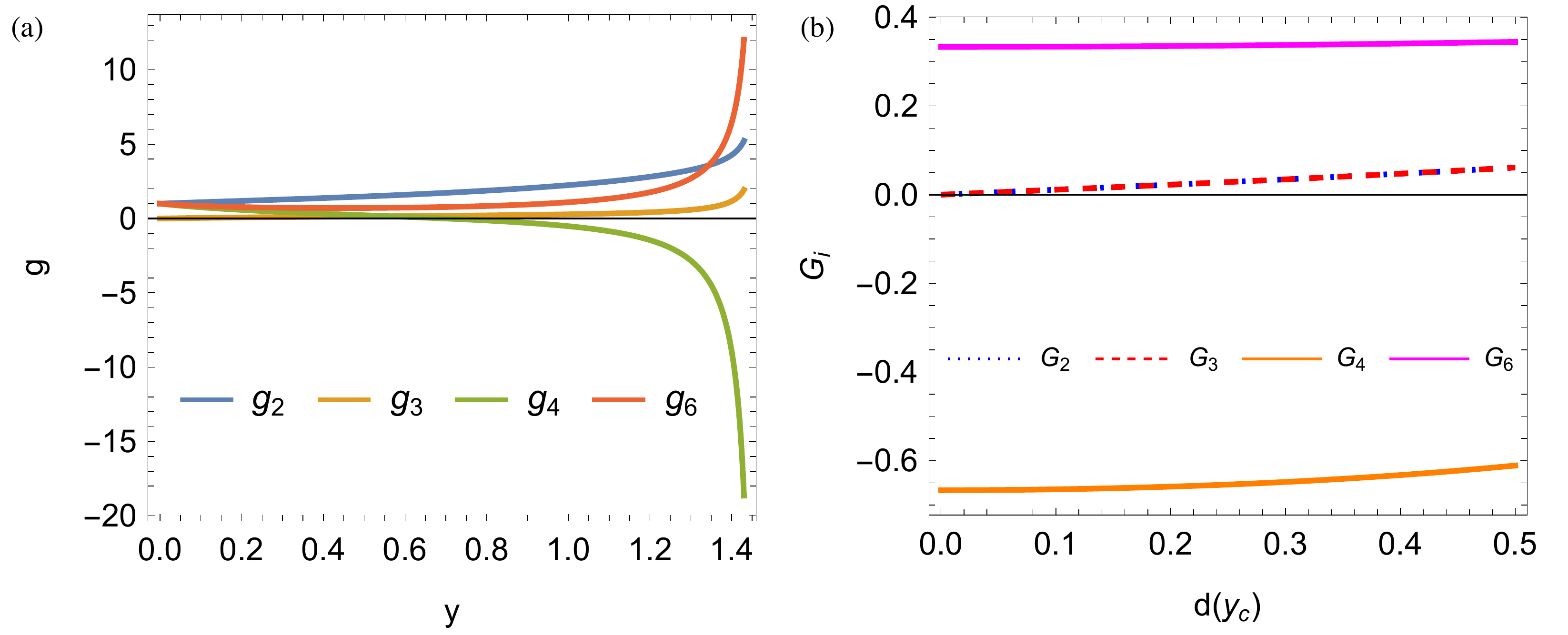}
  \caption{ (a) Numerical solution of \eqref{eq:betaf} with the initial condition $g_2(0)=g_4(0)=g_6(0)=1$. (b) $G_i$'s dependence on $d(y_c)$.} \label{fig:betaf}
\end{figure}
Here we use $y=\Pi_{pp}(0)$ as a running parameter, and $d_1=d\Pi_{ph}(\bm{Q})/dy\approx\Pi_{ph}(\bm{Q})/\Pi_{pp}(0)$, $d_2=d\Pi_{ph}(\bm{Q}')/dy\approx\Pi_{ph}(\bm{Q}')/\Pi_{pp}(0)$. The maximum value of $d_1$ and $d_2$ is $d^{max}=1/2$ as we discussed above. The interactions flows to strong coupling limit at some particular $y_c$, and behave like $g_i=G_i/(y_c-y)$ near this critical value. Substituting this scaling form into the differential equation, we can solve for these $G_i$'s near $y_c$. In fig.\ref{fig:betaf} we show the numerical solution of \eqref{eq:betaf} obtained by modeling $d_1(y)=0.5/\sqrt{1+y}$ and with a particular initial condition $g_2(0)=g_4(0)=g_6(0)=1$ and $g_3(0)=0$ (this is because the initial Hubbard interaction does not contain nonzero $g_3$). We also show the solution of $G_i$'s assuming $d_1(y_c)=d_2(y_c)=d(y_c)$ for different $d(y_c)$ in Fig.\ref{fig:betaf}. It's clear that $G_2$ and $G_3$ are the same, which is only true when $d_1$ and $d_2$ are the same. 

The RG equation for the superconductivity order parameter is shown in Fig.\ref{fig:VHS}(c). Written explicitly, this is
\begin{equation}
  \frac{d}{dy}\begin{pmatrix}
    \Delta_1\\
    \Delta_2\\
    \Delta_3
  \end{pmatrix}=-\begin{pmatrix}
    g_4 ~g_6 ~g_6\\
     g_6 ~g_4 ~g_6\\
      g_6 ~g_6 ~g_4\\
  \end{pmatrix}\begin{pmatrix}
    \Delta_1\\
    \Delta_2\\
    \Delta_3
  \end{pmatrix}\label{eq:beta_SC}
\end{equation}
Diagonalizing the kernel matrix, we see there are two eigenvalues which can be identified as the interaction vertex for the $s/f$-wave SC and the $p/d/g$-wave SC:
\begin{equation}
  \Gamma_{SC}^{s/f}=g_4+2g_6, ~\Gamma_{SC}^{p/d/g}=g_4-g_6.
\end{equation}
Note that since the symmetry is now $D_{3v}$ for nonzero $\phi$, the $p$-wave, $d$-wave and $g$-wave are in the same representation and can not be distinguished, and so are the $s$-wave and $f$-wave.
For the spin-valley density wave order parameters, the RG equation is also shown diagrammatically in Fig.\ref{fig:VHS}(c). For the order parameters peaked at $\bm{Q}$, we have 
\begin{equation}
  \frac{d}{dy}\begin{pmatrix}
    \Delta_{DW1}\\
    \Delta_{DW2}\\
  \end{pmatrix}=d_1\begin{pmatrix}
    g_2 ~g_6\\
     g_6 ~g_2\\
  \end{pmatrix}\begin{pmatrix}
    \Delta_{DW1}\\
    \Delta_{DW2}\\
  \end{pmatrix}\label{eq:beta_DW}
\end{equation}
Again after diagonalization we find the two interaction vertices as
\begin{equation}
  \Gamma_{DW}^+=d_1(g_2+g_6), ~ \Gamma_{DW}^-=d_1(g_2-g_6)
\end{equation}
For the order parameters peaked at $\bm{Q}'$, we have 
\begin{equation}
  \frac{d}{dy}\begin{pmatrix}
    \Delta_{DW'1}\\
    \Delta_{DW'2}\\
  \end{pmatrix}=d_2\begin{pmatrix}
    2g_3 ~-g_6\\
     -g_6 ~2g_3\\
  \end{pmatrix}\begin{pmatrix}
    \Delta_{DW'1}\\
    \Delta_{DW'2}\\
  \end{pmatrix}\label{eq:beta_DWp}
\end{equation}
Again after diagonalization we find the two interaction vertices as
\begin{equation}
  \Gamma_{DW'}^+=d_2(2g_3+g_6), ~ \Gamma_{DW'}^-=d_2(2g_3-g_6)
\end{equation}

The competition among these orders under RG flow can be best seen by looking into the corresponding susceptibilities, which are governed by the following equation:
\begin{equation}
  \frac{d\chi_{SC}}{dy}=\Delta_{SC}^2,~ \frac{d\chi_{DW}}{dy}=d\Delta_{DW}^2,
\end{equation}
Upon integration, they obey the same form $\chi\sim(y_c-y)^\alpha$. If $\alpha<0$ the corresponding susceptibility diverges, signaling an onset of a particular order. The exponents  for the orders are:
\begin{equation}
  \begin{aligned}
    &\alpha_{SC}^{s/f}=2(G_4+2G_6)+1, ~\alpha_{SC}^{p/d/g}=2(G_4-G_6)+1\\
    &\alpha_{DW}^+=-2d_1(G_2+G_6)+1, ~ \alpha_{DW}^-=-2d_1(G_2-G_6)+1,\\
    &\alpha_{DW'}^+=-2d_2(2G_3+G_6)+1, ~ \alpha_{DW}^-=-2d_2(2G_3-G_6)+1,
  \end{aligned}
\end{equation}
Based on these exponents, we find that the $p/d/g$-wave SC is the only possible order as is discussed in the main text.

\label{sub:six_patch_model}


\subsubsection{Two patch model} 
\label{sub:two_patch_model}
\begin{figure}
  \includegraphics[width=14cm]{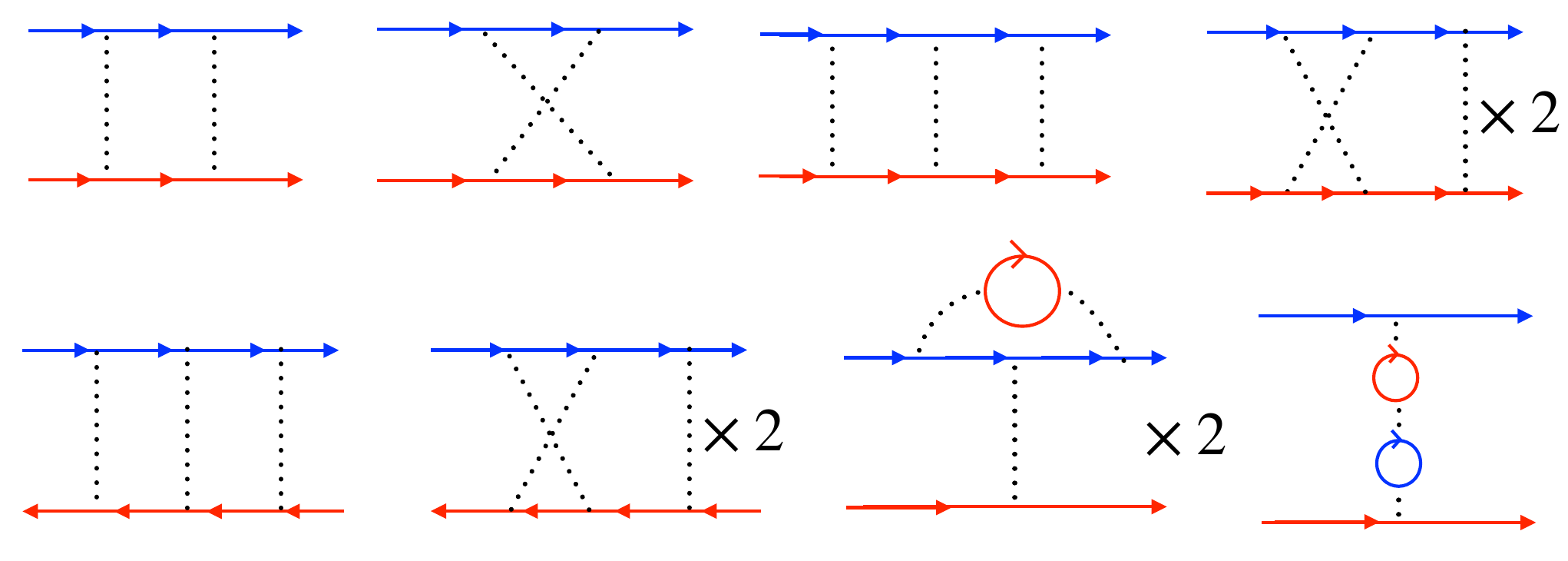}
  \caption{ RG diagrams that contribute to $\dot{g_2}$ in the two patch model up to two-loop level.} \label{fig:2loop}
\end{figure}

In the two patch model we are dealing with two higher order VHSs. In this case, we only need to consider the inter-patch density-density interaction:
\begin{equation}
  H_I=\frac{g_2}{2} \rho_1\rho_2.
\end{equation}
This interaction will renormalize itself once we consider loop corrections. Due to the exact cancellation between the particle-particle bubble at $\bm{Q}=0$ and the particle-hole bubble at $\bm{Q}=\bm{K}$ in this two patch model, there is no one-loop contributions to the renormalization of $g_2$ at perfect nesting. Although the one-loop RG equation becomes nonzero once we consider the system away from perfect nesting, its vanishing at perfect nesting indicates the importance of higher order loop corrections. We thereby obtain the RG equation up to two-loop level:
\begin{equation}
  \dot{g_2}=-(1-d_1)g_2^2+2\left(1+d_1^2-4d_1-\frac{1}{9}d_2^2\right)yg_2^3\label{eq:twoloop}
\end{equation}
The diagrammatic representation of this equation is shown in Fig.\ref{fig:2loop}. Here $y=\Pi_{pp}(0)$ and we have defined $d_1=\Pi_{ph}(\bm{Q})/\Pi_{pp}(0)$ and $d_2=3\Pi_{ph}(0)/\Pi_{pp}(0)$.

When the system is tuned right at the perfect nesting point ($n=1$ and $\phi=\pi/6$), we have $d_1=1$ and $d_2=1$. In this case, the RG equation becomes $\dot{g_2}=-\frac{38}{9}yg_2^3$ and then $g_2$ flows to zero in low temperature limit regardless of sign of its initial value $g_2(0)$. This system remains gapless and develops no order. 

When the system is away from perfect nesting, we have $d_1<1$ and $d_2<1$. The coefficient of $g_2^2$ term in \eqref{eq:twoloop} is finite but negative, while the coefficient of $yg_2^3$ term can be positive or negative, depending on the values of $d_1$ and $d_2$. In the case when this two-loop coefficient becomes positive, $g_2$ flows to strong repulsion at sufficiently low temperature when $g_2(0)$ is above some threshold value. As a result, density-wave order wins over SC in low $T$ limit when $g_2(0)$ is large, but SC is the leading instability in weak coupling limit.

We emphasis that here the case when $d_1<1$ and $d_2<1$, including the limit $d_1\to0$ and $d_2\to0$, is obtained around the higher order VHS, where $\Pi_{ph}(0)$ and $\Pi_{ph}(\bm{Q})$ both diverges as $1/T^{1/3}$. When the system is far away from the higher order VHS, we would expect that the only divergent channel is the Cooper channel, thus SC is the only possible order in low temperature limit.

Nevertheless, we have neglected the tree level scaling dimension of $g_2$ in the gapless regime. This term is possible to change the RG flow qualitatively as $g_2$ is tree level relevant, due to the cubic dispersion of the higher-order Van Hove singularity. Following the Wilsonian energy shell RG scheme for the higher-order Van Hove singularity\cite{Isobe2019}, we further use the Wilsonian RG to include the tree level contribution in the perfect nesting regime: $d_1=1$.

We take the ultraviolet energy cutoff as $\Lambda$ around the higher-order Van Hove singularities, and assume the following tree level scaling dimension to fix the kinetic energy:
\begin{equation}
    \omega\rightarrow (1-dl)\omega,\quad  k_{x,y}\rightarrow (1-dl)^{\frac{1}{3}} k_{x,y},\quad \psi(\Vec{k},\omega)\rightarrow (1-dl)^{-\frac{4}{3}} \psi(\Vec{k},\omega),
\end{equation}
where $\Lambda dl$ is the infinitesimal energy shell. The action up to tree level marginal terms is given by:
\begin{equation}
\begin{aligned}
&S=S_0+S_I,\\
   &S_0=\int d\omega d^2k \bar{\psi}_{\bm{K}}(\Vec{k},\omega)(-i\omega+\epsilon_{\bm{K}}(\Vec{k}))\psi_{\bm{K}}(\Vec{k},\omega)+\bar{\psi}_{-\bm{K}}(\Vec{k},\omega)(-i\omega+\epsilon_{-\bm{K}}(\Vec{k}))\psi_{-\bm{K}}(\Vec{k},\omega),\\
   &S_I=\frac{g_2}{2}\int d\omega d^2k \rho_{\bm{K}}(\Vec{k},\omega) \rho_{-\bm{K}}(-\Vec{k},-\omega).
\end{aligned}
\end{equation}
We should note that there is no relevant or marginal kinetic energy perturbation to convert the higher-order Van Hove singularities into conventional Van Hove singularities due to the $C_3$ rotation symmetry, which is different from that in \cite{Isobe2019}.

We integrate out the fast modes in the energy shell: $[-\Lambda,-\Lambda(1-dl)]\cup [\Lambda(1-dl),\Lambda]$ and arrive at the two loop RG equation for $g_2$ is:
\begin{equation}
    \frac{d \bar{g}_2}{dl}=\frac{1}{3} \bar{g}_2-2\bar{g}_2^3,
\end{equation}
 where $\bar{g}_2=D(\Lambda) g_2$ and $D(\Lambda)$ is the density of states of a single Van Hove singularity at energy $\Lambda$. The Feynman diagrams are still in Fig.\ref{fig:2loop}, except the final two diagrams. These two diagrams vanish due to $\Pi_{ph}(0)=0$ in the energy shell. There is an interacting fixed point at: $\bar{g_2}=\sqrt{\frac{1}{6}}$, which is stable to the perturbation in $\bar{g_2}$: $ \frac{d \delta \bar{g}_2}{dl}=-\frac{2}{3}\delta\bar{g}_2^2$. The filling must be tuned to the Van Hove doping, as the chemical potential is a relevant perturbation up to two loops: $\frac{d\mu}{dl}=\mu$. This interacting fixed point is a non-Fermi liquid dubbed as supermetal in \cite{Isobe2019}, and becomes Fermi liquid when it is doped away from Van Hove singularity.

\subsection{B. Raghu-Kivelson-Scalapino renormalization group analysis for the SC order} 
\label{sec:renormalization_group_analysis}
The specific shape of the FSs for both spin up and spin down fermions depend on the value of $\phi$. For the regular case free of van Hove singularities, there is no nesting effect along the FS.
 As a result, we can consider only SC instabilities. Since we are interested in the possible pairing symmetry, we need to take all momenta on the FS into account. 

\begin{figure}
  \includegraphics[width=12cm]{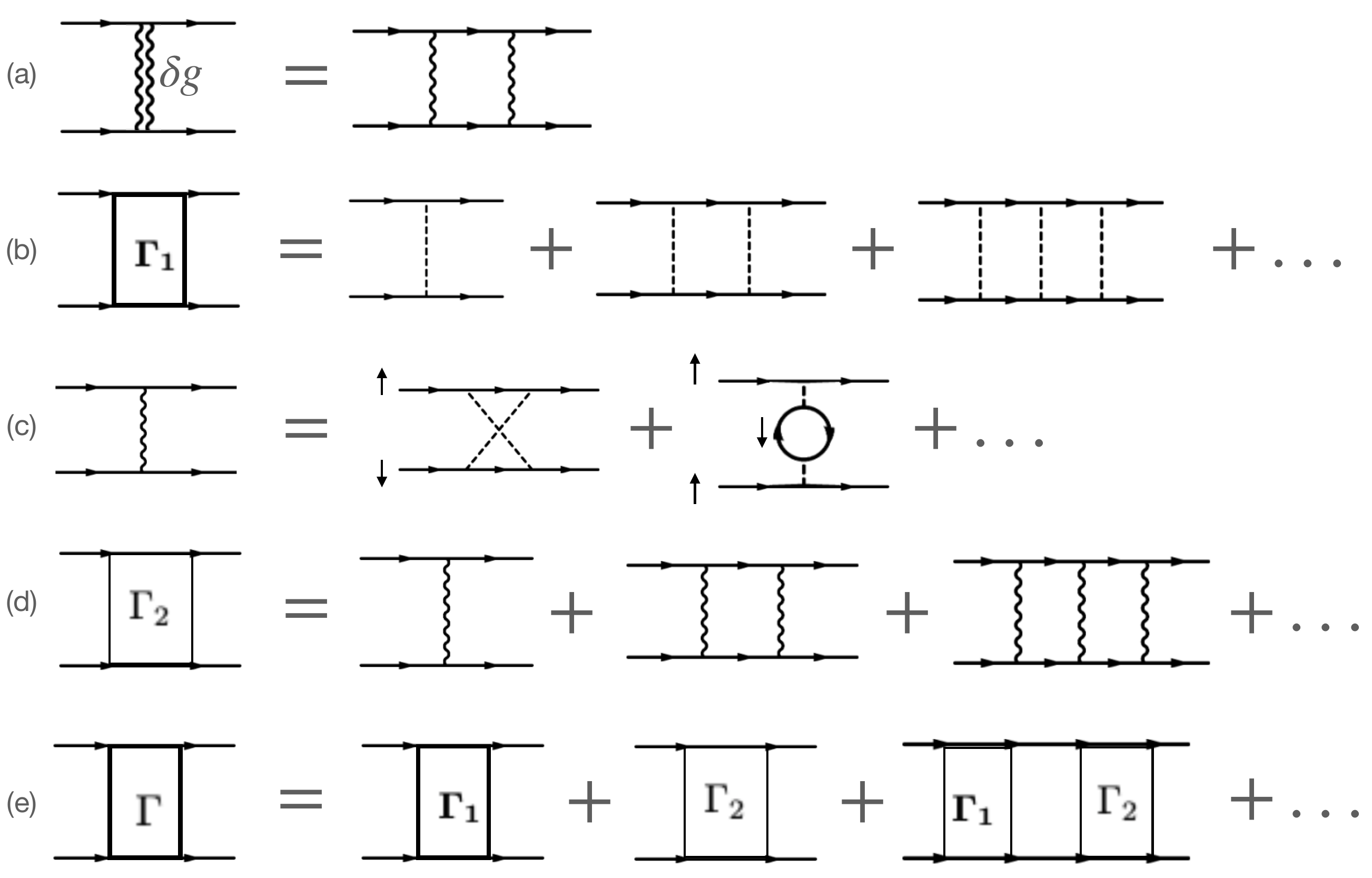}
  \caption{Diagrammatic interpretation of the renormalization group analysis.}\label{fig:diagram}
\end{figure}

The possible SC from a repulsive Hubbard $U$ is a higher order effect. This should be in contrast to the attractive $U$ case, where an arbitrarily small attractive $U$ could lead to Cooper instability. To see this, it is convenient to consider the one-loop renormalization group equation for the interaction vertex in Cooper channel:
\begin{equation}
  \frac{dg}{dl}=-g\star g,\label{eq:RG}
\end{equation}
which is also shown pictorially in Fig. \ref{fig:diagram}(a). Here if we adopt finite temperature RG scheme, $l=\rho\log{W/T}$ with $\rho$ being density of states and $W$ the band width, is the Cooper logarithm from the particle-particle bubble and plays the role of the running parameter. In zero temperature RG scheme, the running parameter is $l=\rho\log W/\Lambda$ where $\Lambda$ is the running energy scale which changes as one continuously integrates out fast fermion modes. The transition temperature $T_c$ and the zero-$T$ gap amplitudes are the same with the energy scale $\Lambda^*$ when the RG procedure breaks down.
Here $g$ is the interaction vertex, which can be bare or dressed. The notation $\star$ is a shorthand of the integral convolution between the two involved vertices, but for the simplest case when $g$ is the bare interaction $U$, it is equivalent to scalar multiplication.

Let's first consider the case when $g=\Gamma=U$. Solving the differential equation with the initial condition $\Gamma(0)=U$, one obtains
\begin{equation}
  \Gamma(l)=\frac{1}{l+1/U}=\frac{U}{1+lU}.\label{eq:Gammal}
\end{equation}
Notice that exactly the same expression can also be obtained by summing up the infinite series of the ladder diagrams shown in Fig.~\ref{fig:diagram}(b). This has demonstrated the equivalence between the one-loop RG analysis and the diagrammatic calculation. Now it becomes manifest that in the low energy limit, the interaction either flows to zero, or enter strong coupling regime, depending on whether $U>0$ or $U<0$. Indeed, for any arbitrarily small attractive $U$, the point when $1+lU=0$ is where RG fails to work, and this is the Cooper instability which marks the onset of superconductivity. The relevant temperature can then be found readily from this condition, which is $T_c\sim \exp[-1/(\rho|U|)]$. For a positive $U$, however, there is no such an instability at the bare interaction level. 

SC could arise for $U>0$ if we consider, instead of the bare $U$, the dressed {\it static} interaction vertex shown in Fig. \ref{fig:diagram}(c), and use it as the building block in the ladder series depicted in Fig. \ref{fig:diagram}(d). Again, this ladder series is an equivalent way to interpret the one-loop RG equation in \eqref{eq:RG}. The dressed interaction vertices in Fig. \ref{fig:diagram}(c) are those which do not contain Cooper logarithm, and we only show the diagrams up to the second order in the Hubbard $U$. For weak interacting system which we consider in this work, these two are the leading order corrections to the bare $U$. The full interaction vertex in the pairing channel, contains contributions from both $\Gamma_1$ and $\Gamma_2$, as is shown in Fig. \ref{fig:diagram}(e). It is easy to see the the combination of $\Gamma_1$ with another Cooper logarithm gives a regular contribution, and as a result, the leading possible instability now is solely determined by $\Gamma_2$, which is a function of the external momenta.  Unlike the case with bare interaction, the resulting pairing symmetry is usually no longer $s$-wave.  For practical purpose, one should also distinguish different contributions in $\Gamma_2$ to pairing states with different spin configurations. In the Hubbard model, the bare interaction $U$ connects fermions with opposite spin projections. Consequently, the first diagram in Fig. \ref{fig:diagram}(c) contributes to $S_z=0$ pairing state, while the second one contributes to $|S_z|=1$ state.

For $\phi\neq0$, fermions with different spins have different energy dispersions. What's more, the Fermi surface of the spin up or down electrons are not exactly inversion symmetric around the $\Gamma$ point. As a result, the $|S_z|=1$ pairing is less favourable than the  $|S_z|=0$ pairing with a generic nonzero $\phi$. Hence, we focus on  $|S_z|=0$ pairing with a general $\phi\neq 0$ hereafter. 
There is one exception when $\phi$ can be related to $\phi=0$ via symmetry transformation. In this case the competition between a finite $\bm{Q}$ PDW in $|S_z|=1$ channel and a $\bm{Q}=0$ SC state in $S_z=0$ channel becomes possible. Thus we will be focusing only on the $S_z=0$ channel when $\phi$ is neither zero nor in other high symmetry point. In the special case of $\phi=0,\pi/3$, pairing with same spin becomes possible, and accordingly the second diagram in Fig. \ref{fig:diagram}(c) needs to be considered together with the first one, leading to a competition between singlet and triplet pairing.

Based on the above discussion the resulting interaction vertex to be input into \eqref{eq:RG} can be written as \cite{Raghu2010,Cho2013}
\begin{equation}
  g_{k,p}=\sqrt{\frac{\bar{v}_F}{|v_F(k)|}}\Gamma(k,p)\sqrt{\frac{\bar{v}_F}{|v_F(p)|}},\label{eq:gkp_s}
\end{equation}
We are at liberty to choose both $k$ and $p$ on, e.g. the spin up FS, and the Fermi velocities then must be calculated from this particular choice of FS. One is also free to flip the sign of one or both of these external momenta, but it changes nothing since the FS is fixed as we fix the spin projection, and $k$ and $-k$ are just two different notations for the same FS. The interaction vertex for $\phi\neq 0,\pi/3$, to the order of $U^2$, is just the static particle-hole bubble:
\begin{equation}
  \Gamma(\bm{k},\bm{p})= U^2\chi_{S_z=0}(\bm{k}+\bm{p}).
\end{equation}
For $\phi=0$ and $\phi=\pi/3$ we need to replace $\chi_{S_z=0}(\bm{k}+\bm{p})$ in the above expression with $-\chi_{|S_z|=1}(\bm{k}-\bm{p})$, where the minus sign comes from the fermion loop in Fig.~\ref{fig:diagram}(c). For our model, the particle-hole bubbles in different channels are:
\begin{equation}
  \begin{aligned}
    \chi_{S_z=0}(\bm{p}+\bm{k})&=\int_{BZ}\frac{d^2q}{8\pi^2}\frac{\tanh[\beta \xi_{\bm{p}+\bm{k}+\bm{q}}^+/2]-\tanh[\beta \xi_{\bm{q}}^-/2]}{\xi_{\bm{p}+\bm{k}+\bm{q}}^+-\xi_{\bm{q}}^-},\\
    \chi_{S_z=1}(\bm{p}-\bm{k})&=\int_{BZ}\frac{d^2q}{8\pi^2}\frac{\tanh[\beta \xi_{\bm{p}-\bm{k}+\bm{q}}^+/2]-\tanh[\beta \xi_{\bm{q}}^+/2]}{\xi_{\bm{p}-\bm{k}+\bm{q}}^+-\xi_{\bm{q}}^+}.
  \end{aligned}\label{eq:numericalChi}
\end{equation}
Here $\beta$ is the inverse temperature which can be taken as infinitely large in zero temperature limit, $\xi=\epsilon-\mu$ is from the energy dispersion discussed in our main text.

 The additional factor $\sqrt{\bar{v}_F/|v_F(k)|}$ in \eqref{eq:gkp_s} comes from the need of the evaluation of 
\begin{equation}
  \int\frac{d^2k}{(2\pi)^2}\Gamma(\bm{k},\bm{p})\left[\frac{1}{\beta}\sum_n G(k,\omega_n)G(-k,-\omega_n)\right]\label{eq:Gammavertex}
\end{equation}
when summing up the series in Fig\ref{fig:diagram}.(d). The particle-particle and particle-hole bubbles are convoluted, but in low energy limit can be separated by considering a thin shell around the FS where integration over momentum can be carried out along two directions: along and perpendicular to the FS. In the perpendicular direction $\chi(k+p)$ is treated as a constant, and the integration leads to Cooper logarithm. To see this it's convenient to re-express the integration as follows,
\begin{equation}
  \int d k_{\bot}=\frac{(2\pi)^2}{L_F}\frac{\bar{v}_F}{|v_F(k)|}\int\rho d\varepsilon,
\end{equation}
where  $L_F$ is the length of the FS (the FS area in 2D). The average Fermi velocity on FS is obtained via
\begin{equation}
  \frac{1}{\bar{v}_F}=\int\frac{dk_{\Vert}}{L_F}\frac{1}{|v_F(k)|}.
\end{equation}
The remaining integration along FS thus incorporates the factor $\bar{v}_F/|v_F(k)|$. Distribute it symmetrically to the two adjacent $\Gamma_{S_z}$ connected by a particle-particle bubble, we immediately arrive at \eqref{eq:gkp_s}. The $\star$ operator in \eqref{eq:RG} can also be easily deduced from the remaining integration on FS. Written explicitly we have
\begin{equation}
  (g\star g)_{k,p}=\int\frac{dk'_\Vert}{L_F}g_{k,k'}g_{k',p}.
\end{equation}
For a set of discrete momentum points, this is nothing but a matrix product, accompanied by a constant factor $\delta k'_\Vert/L_F$.

To solve the RG equation \eqref{eq:RG} where $g$ is in fact a matrix with infinite dimension in the continuum limit, it is easy to first diagonalize it since $g$ is a real symmetric matrix. The eigenvalues $\lambda_n$ from diagonalization are all decoupled in the RG equation, and each of them is constrained by the same equation $d\lambda_n/dl=-\lambda_n^2$. The solution for each $\lambda(l)$ obeys exactly the same form as \eqref{eq:Gammal}. From this we can easily determine whether or not the system will develop a SC instability. If one or a few eigenvalues are negative this channel will flow to strong coupling regime in low energy limit, then indeed there will be SC phase transition in low energy limit. The smallest negative eigenvalues corresponds to the SC state with the largest transition temperature. In general, different eigenvalues corresponds to different pairing symmetry. To appreciate this, consider the linearized gap equation from the dressed interaction:
\begin{equation}
  \Delta(p)=-\int\frac{d^2k}{(2\pi)^2}\Gamma(p,k)\left[\frac{1}{\beta}\sum_n G(k,\omega_n)G(-k,-\omega_n)\right]\Delta(k).
\end{equation}
Using the same integration tactic discussed above, we can easily see that the eigenvectors $g_{p,k}$ are the gap functions up to some nonzero factor $\tilde{\Delta}(p)=\sqrt{\bar{v}_F/|v_F(p)|}\Delta(p)$. Therefore, the pairing symmetry is determined by the structure of the eigenfunction from diagonalization of $g$.

\begin{figure}
  \includegraphics[width=8cm]{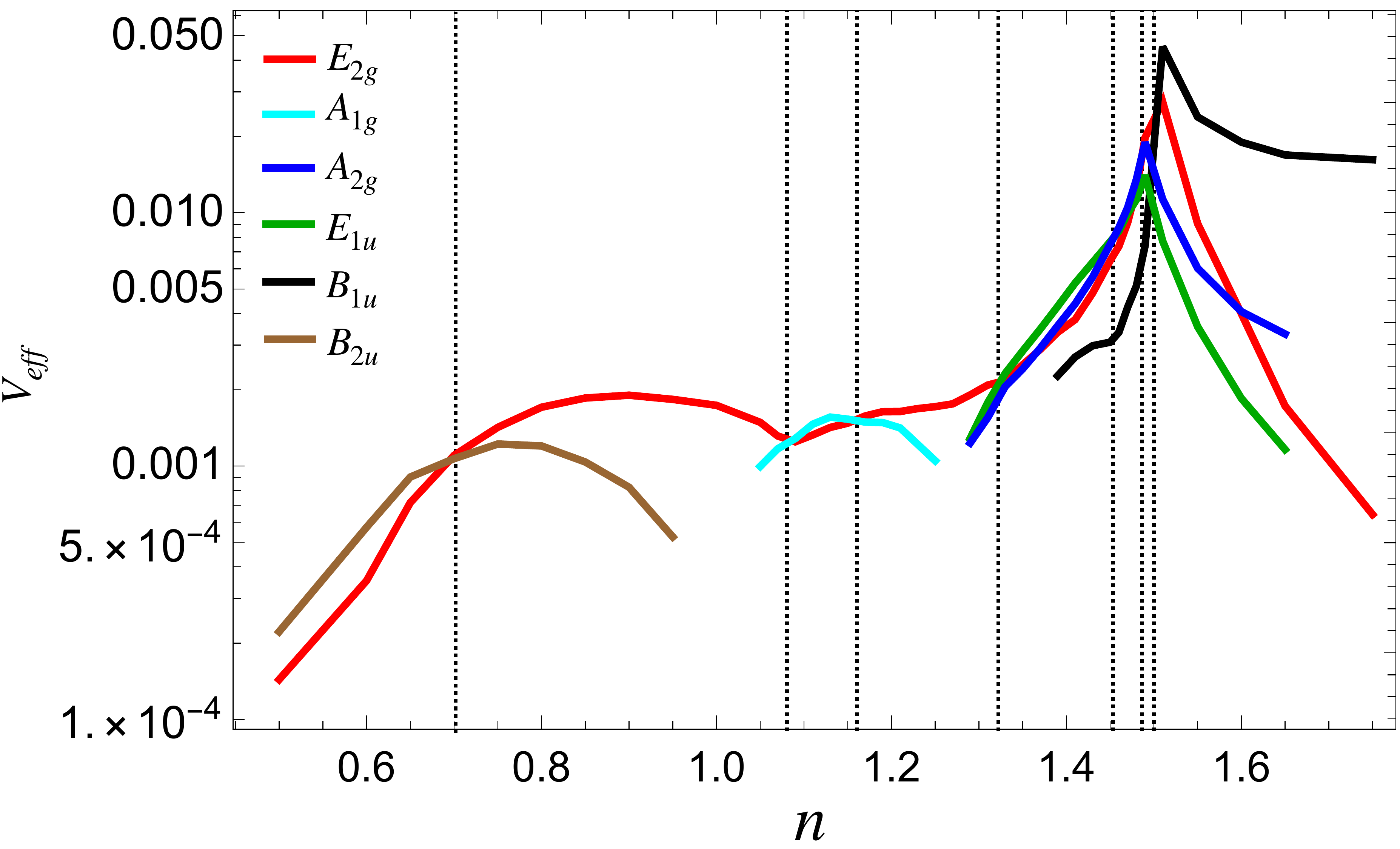}
  \includegraphics[width=8.5cm]{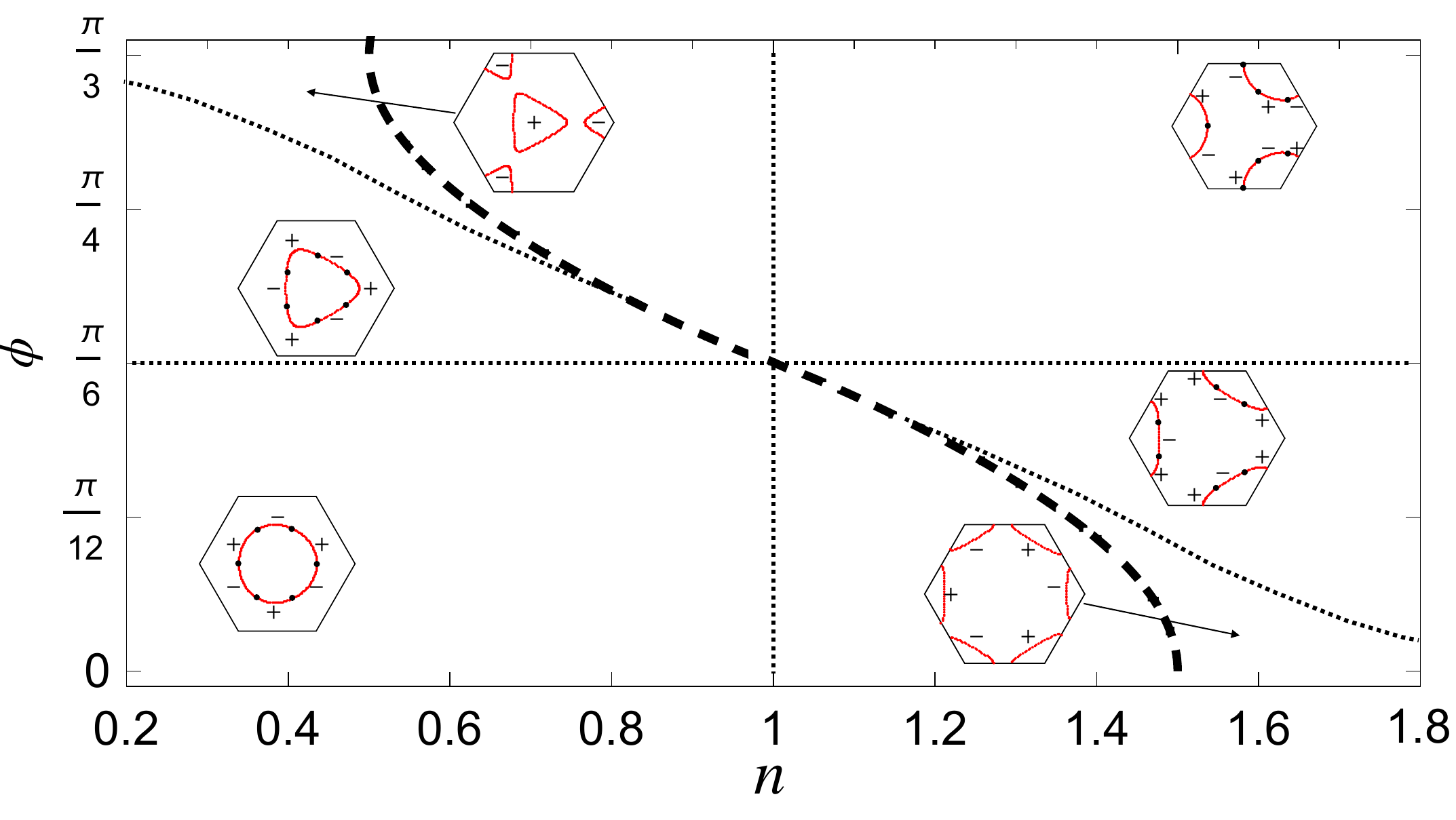}
  \caption{Left: Pairing strength in different channels at $\phi=0$. Right: Nodal and nodeless gap functions belonging to $A_1$ and $A_2$ representations at small and large $n$ with $\phi\neq0$. }\label{fig:phizero}
\end{figure}

As an example, we plot in Fig.~\ref{fig:phizero} the pairing strength $V_{eff}$ in different channels at $\phi=0$ when the symmetry is $D_{6h}$.  To obtain this and other results in our work, we cut the Fermi surfaces into N equal-length segments. For systems away from VHS, we use $N=100$, for systems close to a VHS, we increasse $N$ to around $300$.  The dimensionless pairing strength is defined through the eigenvalue $V_{eff}=\rho|\lambda|$ such that the SC transition temperature scales as $\exp[-1/(V_{eff}U^2/t^2)]$. In the same figure, we also show some typical examples of the gap function when $\phi\neq0$ and $n$ is small and large. Under this circumstance, the ground state is either a six node or nodeless SC state, which dictates the $A_1$ and $A_2$ representation of the underlying $C_{3v}$ group. Note that $A_1$ representation is symmetric while $A_2$ representation is anti-symmetric under $\sigma_v$ operation. For a fixed $n$ (say, $n=1.7$), as $\phi$ increases, the gap function undergoes a transition from $A_1$ representation to $A_2$ representation. The transition is not sharp as there is a finite range around $\phi=\pi/6$ where the pairing strength of the two representations are very comparable.

\subsection{C. Ginzburg-Landau free energy} 
\label{sec:ginzburg_landau_free_energy}
The true ground state should minimize the Ginzburg-Landau free energy, and this criterion is used to determine the two complex amplitudes $\Delta_1$ and $\Delta_2$. To this end, we start with the action in pairing channel, then perform Hubbard-Stratonavich transformation by introducing the gap function $\Delta(\bm{k})$ as an  auxiliary field. The resulting Lagrangian is
\begin{equation}
  \begin{aligned}
    \mathcal{L}=&-\sum_{\bm{p},\bm{k}}\bar{\Delta}(\bm{p})g_{\bm{p},\bm{k}}^{-1}\Delta(\bm{k})\\
    &-\tr_\bot\left[ \Psi^\dagger(\bm{k})
  \begin{pmatrix}
    G_+^{-1} & \Delta_d(\bm{k})\\
    \bar{\Delta}_d(\bm{k}) & G_-^{-1}
  \end{pmatrix}
  \Psi(\bm{k})\right].\label{eq:lag1}
  \end{aligned}
\end{equation}
Here $\Delta(\bm{k})$ should be understood as a column vector in momentum space with $\bm{k}\in\text{FS}$. In the second line $\Psi(\bm{k})=[c_{\bm{k}_1},c_{\bm{k}_2},...,c^\dagger_{-\bm{k}_1},c^\dagger_{-\bm{k}_2},...]^T$ is a generalized Nambu-Gorkov basis with all $\bm{k}_i\in\text{FS}$. Therefore the off-diagonal term $\Delta_d(\bm{k})=\Delta_1d_1+\Delta_2d_2$ is a diagonal matrix with $d_1=\text{diag}[v_1(\bm{k})]$ and $d_2=\text{diag}[v_2(\bm{k})]$, where $v_1$ and $v_2$ are the two degenerate eigenstates of the $g$-matrix. The diagonal terms in \eqref{eq:lag1} are the particle and hole Green's functions defined by
\begin{equation}
  G_\pm=\frac{1}{i\omega_m\mp\xi_{\pm\bm{k}}^{\pm}}.
\end{equation}
Unlike in $\Psi$ and $\Delta_d$, the momentum $\bm{k}$ in $G_\pm$ needn't to be restricted on FS. But all momenta have to be integrated out. We again use the same strategy applied to \eqref{eq:Gammavertex} already discussed above. The spirit is to approximately separate the integration in two orthogonal directions. Here we already adopted this in \eqref{eq:lag1}, noting that while the matrix product in the second line of \eqref{eq:lag1} is in fact a convenient way of expressing the integration along the FS, the partial trace $\tr_\bot$ takes care of the integration perpendicular to the FS as well as dimensional factors.

If we substitute $\Delta(\bm{k})$ from the combination of the two degenerate eigenvectors of $g$ into the Lagrangian,  then  the first term can be simplified using the orthogonality between these two eigenvectors. To obtain the effective Ginzburt-Landau free energy, we need to further integrate out fermion degrees of freedom $\Psi$, and expand the action to the order of $O(\Delta^4)$. In the intermediate stage one encounters a quadratic term $\tr_\Vert(G_+G_-\Delta_d\Delta_d^*)$ as well as a quartic term $\tr_{\Vert}(G_+G_-G_+G_-\Delta_d\Delta_d^*\Delta_d\Delta_d^*)$ where $\tr_\Vert$ denotes the integration along the FS. Given that $G_\pm$ does not depend on $\bm{k}\in\text{FS}$, the Green's functions can be pulled out of the $\tr_\Vert$ operation. Thus the partial trace operation to the FS acts only on the gap function and it can be further simplified using the fact that $\tr_\Vert(d_1^2)=\tr_\Vert(d_2^2)=1$, $\tr_\Vert(d_1^4)=\tr_\Vert(d_2^4)=3\tr_\Vert(d_1^2d_2^2)$ while other cross terms such as $\tr_\Vert(d_1d_2)$ and $\tr_\Vert(d_1^3d_2)$ all vanish. All these considerations lead us to the following free energy:
\begin{equation}
  \begin{aligned}
    \mathcal{F}[\Delta_1,\Delta_2]&=\alpha(T-T_c)(|\Delta_1|^2+|\Delta_2|^2)\\
    &+\beta_1(|\Delta_1|^2+|\Delta_2|^2)^2+\beta_2|\Delta_1^2+\Delta_2^2|^2\label{eq:F2}
  \end{aligned}
\end{equation}
where $\alpha(T-T_c)=\tr_\bot(G_+G_-)-\lambda_{min}^{-1}$ with $\lambda_{min}<0$ being the smallest degenerate eigenvalue of the $g$-matrix, and $\beta_1=(1/3)K\tr_\Vert(d_1^4)$ and $\beta_2=(1/2)K\tr_\Vert(d_1^2d_2^2)$ with $K=\tr_\bot(G_+G_-G_+G_-)$. Note that $\tr_\bot(G_+G_-)=-\rho\ln[W/T]$ is the Cooper logarithm from which we can determine $T_c$, and the quartic term $K$ is always positive.

\end{document}